\documentclass[11pt]{article}
\usepackage{url}

\usepackage{hyperref}

\usepackage{amssymb,amscd,amstext,amsmath, amsthm, graphicx}
\usepackage{latexsym}
\usepackage[russian,english]{babel}
\usepackage{floatflt}
\usepackage[mathscr]{eucal}
\usepackage{tikz}
\usetikzlibrary{arrows}
\usepackage{verbatim}
\usepackage{soul}
\usepackage{wrapfig}
\usepackage{fancyhdr}
\usepackage{hyperref}
\usepackage{setspace}
\usepackage{mathdots}
\usepackage{blindtext}
\usepackage{enumitem}
 \usepackage{placeins}
 \usepackage{algorithm}
 \usepackage{csquotes}
\usepackage{multirow}
\usepackage{longtable}

\usepackage[top=1in, bottom=1in, left=1in, right=1in]{geometry}

\usepackage{tikz}\usetikzlibrary{shadows}

\definecolor{linkcolor}{HTML}{000000}
\definecolor{urlcolor}{HTML}{000000}
\hypersetup{pdfstartview=Fit, linkcolor=linkcolor,urlcolor=urlcolor,
colorlinks=true, citecolor=linkcolor}
\renewenvironment{abstract}{%
\begin{center}\begin{minipage}{0.9\textwidth}\begin{small}
\textbf{Abstract.}}
{\end{small}\par\noindent\end{minipage}\end{center}}

\title{\bf The Seventh International Olympiad in Cryptography: problems and solutions\footnote{The work of the second and sixth authors was supported by Mathematical Center in Akademgorodok under agreement No. 075-15-2019-1613 with the Ministry of Science and Higher Education of the Russian Federation, by HUAWEI CLOUD, and by Laboratory of Cryptography JetBrains Research. The work of the first author was carried out within the framework of the state contract of the Sobolev Institute of Mathematics (project no. FWNF-2022-0018). The work of the first, fifth and eighth authors was supported by Russian Foundation for Basic Research (project no. 20-31-70043).}
}

\author{A.~Gorodilova$^{1}$,
    N.~Tokareva$^{1,2}$,
    S.~Agievich$^{3}$,
    C.~Carlet$^{4}$,
    V.~Idrisova$^{1}$,
    K.~Kalgin$^{1,5}$,\\
    D.~Kolegov$^{6}$,
    A.~Kutsenko$^{1,5}$,
    N.~Mouha$^{7}$,
    M.~Pudovkina$^{8}$,
    A.~Udovenko$^{9}$
    \\
  \\
  {\small$^{1}$Sobolev Institute of Mathematics, Novosibirsk, Russia} \\
  {\small$^{2}$Laboratory of Cryptography JetBrains Research, Novosibirsk, Russia} \\
  {\small$^{3}$Belarusian State University, Minsk, Belarus} \\
  {\small$^{4}$University of Paris 8, Paris, France} \\
  {\small$^{5}$Novosibirsk State University, Novosibirsk, Russia}\\
  {\small$^{6}$Tomsk State University, Tomsk, Russia}\\
  {\small$^{7}$National Institute of Standards and Technology, Gaithersburg, United States}\\
  {\small$^{8}$Bauman Moscow State Technical University, Moscow, Russia}\\
  {\small$^{9}$CryptoExperts, Paris, France}\\
    \\
    {\small E-mail: {\tt nsucrypto@nsu.ru}}
    }

\date{}

\begin{document}

\hypersetup{pageanchor=false}

\begin{titlepage}
\maketitle
\begin{abstract}
The International Olympiad in Cryptography NSUCRYPTO is the unique olympiad containing scientific mathematical problems for professionals, school and university students from any country. Its aim is to involve young researchers in solving curious and tough scientific problems of modern cryptography. In 2020, it was held for the seventh time. Prizes and diplomas were awarded to 84 participants in the first round and 49 teams in the second round from 32 countries. In this paper, problems and their solutions of NSUCRYPTO'2020 are presented. We consider problems related to attacks on ciphers and hash functions, protocols, permutations, primality tests, etc. We discuss several open problems on JPEG encoding, Miller --- Rabin primality test, special bases in the vector space, AES-GCM. The problem of a modified Miller --- Rabin primality test was solved during the Olympiad. The problem for finding special bases was partially solved.

\vspace{0.2cm}

\end{abstract}
\end{titlepage}

\hypersetup{pageanchor=true}
\pagenumbering{arabic}

\setcounter{page}{2}

\section{Introduction}

\begin{figure}[!h]
\centering
\includegraphics[width=0.9\textwidth]{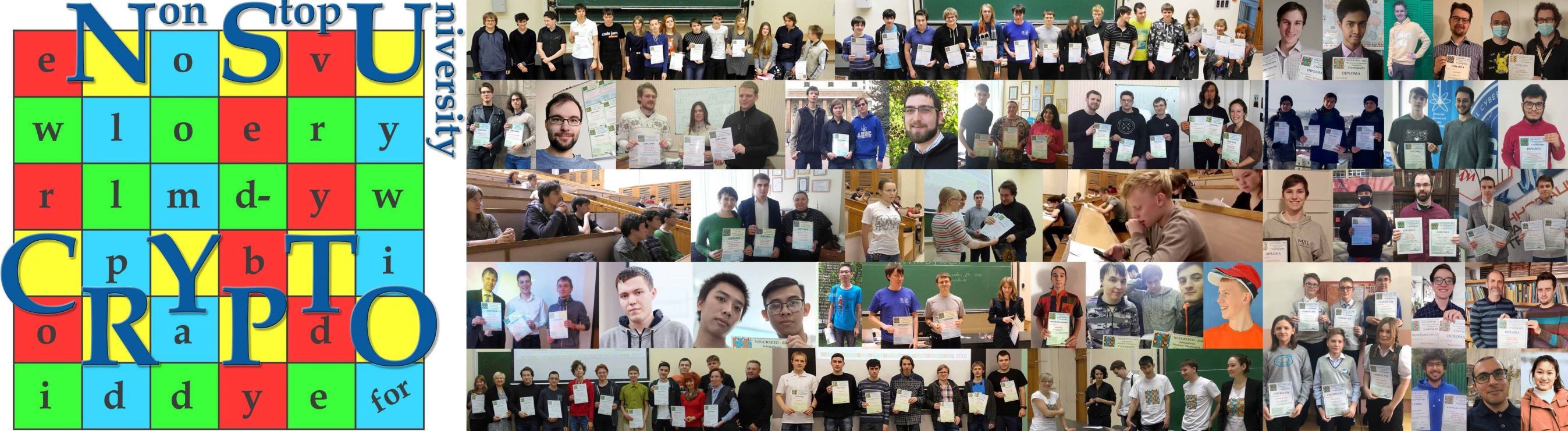}
\vspace{-3mm}
\caption{{\small NSUCRYPTO logo and winners}}
\label{fig:logo}
\end{figure}

NSUCRYPTO (Non-Stop University Crypto) is the International Olympiad in Cryptography that was held for the seventh time in 2020. The Olympiad program committee includes specialists from Belgium, France, the Netherlands, the USA, Norway, India, Luxembourg, Belarus', Kazakhstan, and Russia.
Interest in the Olympiad around the world becomes more significant. In 2020, there were 775 participants from more than 50 countries; and 14 countries took part for the first time. Summing the results, 84 participants in the first round and 49 teams in the second round from 32 countries were awarded with prizes and honorable diplomas. The list of the winners can be found at the official \href{https://nsucrypto.nsu.ru/}{\textcolor{blue}{website}} of the Olympiad \cite{nsucrypto}. Fig.~\ref{fig:logo} illustrates the Olympiad logo and winners.

Let us shortly formulate the format of the Olympiad. When registering to the Olympiad, each participant chooses his/her category: ``school students'' (for junior researchers: pupils and high school students), ``university students'' (for participants who are currently studying at universities) and ``professionals'' (for participants who have already completed education or just want to be in the restriction-free category).
The Olympiad consists of two independent the Internet rounds. The first round is individual (duration 4 hours 30 minutes, two sections: A is for ``school students'', B is for ``university students'' and ``professionals''). The second round is a team one (duration 1 week, common to all participants).

A distinctive feature of the Olympiad is that some unsolved problems at the intersection of mathematics and cryptography are offered to the participants as well as problems with known solutions.
During the Olympiad, one of such open problems, ``Miller --- Rabin revisited'' (see section~\ref{miller-rabin}), was solved completely. For another one problem, ``Bases'' (see section~\ref{bases}), a partial solution was proposed. All the open problems stated during the Olympiad history can be found \href{https://nsucrypto.nsu.ru/unsolved-problems}{\textcolor{blue}{here}} \cite{nsucrypto-unsolved}. What is more important for us that some researchers were trying to find solutions after the Olympiad was over. In the recent paper \cite{20-Kiss-Nagy}, a complete solution was found for the problem ``Orthogonal arrays'' (2018).  A partial solution for the problem ``A secret sharing'' (2014) was proposed in \cite{sharing}.
We invite everybody who has ideas on how to solve the problems to send your solutions to~us!

We start with problem structure of the Olympiad in section~\ref{problem-structure}. Then we present formulations of all the problems stated during the Olympiad and give their detailed solutions in section~\ref{problems}.
Mathematical problems and their solutions of the previous International Olympiads in cryptography NSUCRYPTO from 2014 to 2019 can be found in \cite{nsucrypto-2014}, \cite{nsucrypto-2015}, \cite{nsucrypto-2016}, \cite{nsucrypto-2017}, \cite{nsucrypto-2018}, and \cite{nsucrypto-2019} respectively.

\section{Problem structure of the Olympiad}
\label{problem-structure}

There were 14 problems stated during the Olympiad, some of them were included in both rounds (Tables\;\ref{Probl-First},\,\ref{Probl-Second}). Section A of the first round consisted of six problems, whereas the section B contained seven problems. The second round was composed of ten problems. Four problems included unsolved questions (awarded special prizes from the Program Committee).

\begin{table}[ht]
\centering\footnotesize
\caption{{\bf Problems of the first round}}
\medskip
\label{Probl-First}
\begin{tabular}{cc}
\begin{tabular}{|c|l|c|}
  \hline
  N & Problem title & Max score \\
  \hline
  \hline
  1 & \hyperlink{pr-2020}{2020} & 4 \\
    \hline
  2 & \hyperlink{pr-poly}{{\tt POLY}}  & 4 \\
    \hline
  3 & \hyperlink{pr-house}{A secret house} & 4 \\
    \hline
  4 &  \hyperlink{pr-RGB}{RGB} & 4 \\
    \hline
  5 & \hyperlink{pr-miller-rabin}{Miller --- Rabin revisited (Q1)} & 4 \\
    \hline
  6 & \hyperlink{pr-event}{Mysterious event} & 4 \\
  \hline
\end{tabular}
&
\begin{tabular}{|c|l|c|}
  \hline
  N & Problem title & Max score \\
  \hline
    \hline
  1 & \hyperlink{pr-2020}{2020}  & 4 \\
    \hline
  2 & \hyperlink{pr-house}{A secret house} & 4 \\
    \hline
  3 &  \hyperlink{pr-miller-rabin}{Miller --- Rabin revisited} & 4 + add.  \\
    \hline
  4 & \hyperlink{pr-RGB}{RGB}  & 4 \\
    \hline
  5 & \hyperlink{pr-event}{Mysterious event}  & 4 \\
    \hline
  6 & \hyperlink{pr-cpa}{CPA game} & 6 \\
  \hline
  7 & \hyperlink{pr-collisions}{Collisions (Q1)}  & 4 \\
  \hline
\end{tabular}
\\
\noalign{\smallskip}
Section A
&
Section B
\\
\end{tabular}
\end{table}

\vspace{-0.7cm}

\begin{table}[ht]
\centering\footnotesize
\caption{{\bf Problems of the second round}}
\medskip
\label{Probl-Second}
\begin{tabular}{|c|l|c|}
  \hline
  N & Problem title & Maximum score \\
  \hline
    \hline
  1 &  \hyperlink{pr-poly}{{\tt POLY}}  & 4 \\
    \hline
  2 &  \hyperlink{pr-stairs}{Stairs-Box} & 7 \\
    \hline
  3 &  \hyperlink{pr-RSA}{Hidden RSA}  & 6 \\
    \hline
  4 &  \hyperlink{pr-ortho}{Orthomorphisms}  & 12 \\
    \hline
  5 &  \hyperlink{pr-jpeg}{JPEG Encoding}  & Unlimited (open problem) \\
    \hline
  6 & \hyperlink{pr-miller-rabin}{Miller --- Rabin revisited} & 4 + add. sc. for open pr. \\
    \hline
  7 &  \hyperlink{pr-cpa}{CPA game} & 6 \\
    \hline
  8 &  \hyperlink{pr-collisions}{Collisions} & 8 \\
    \hline
  9 & \hyperlink{pr-bases}{Bases}  & Unlimited (open problem) \\
    \hline
  10 & \hyperlink{pr-aes}{AES-GCM} & 10 + add. sc. for open pr. \\
    \hline
\end{tabular}
\end{table}

\section{Problems and their solutions}\label{problems}

In this section, we formulate all the problems of NSUCRYPTO'2020 and present their detailed solutions paying attention to solutions proposed by the participants.

\subsection{Problem ``2020''}

\subsubsection{Formulation}
\hypertarget{pr-2020}{}

A cipher machine {\tt WINSTON} can transform a binary sequence in the following way. A sequence $S$ is given, a cipher machine can add to $S$ or remove from $S$ any subsequence of the form $11$, $101$, $1001$, $10\ldots01$. Also, it can add to $S$ or remove from $S$ any number of zeros.

When special agent Smith entered the room there were two identical {\tt WINSTON} machines. He was curious to encrypt number 2020 and he tried to encrypt the number in it's binary form. The first cipher machine returned the binary form of number 1984, the second one returned the binary form of number 2021. Smith understood that one of the machines is broken. How did he know that?

\subsubsection{Solution}

By removing subsequences of the form 10...01 and 0...0, the parity of ones in the binary representation cannot be changed. The given numbers have the following binary representations:
\begin{align*}
2020 \rightarrow 11111100100 \rightarrow 7 \text{ ones,}\\
2021 \rightarrow 11111100101 \rightarrow 8 \text{ ones,}\\
1984 \rightarrow 11111000000 \rightarrow 5 \text{ ones.}
\end{align*}

Hence, it is impossible to obtain 2021 from the input 2020. Hence, the second machine must be broken.

\subsection{Problem ``{\tt POLY}''}
\hypertarget{pr-poly}{}
\subsubsection{Formulation}

During a job interview, Bob was proposed to think up a small cryptosystem that operates with integers. Bob invented and implemented a complex algorithm {\tt POLY} that can be represented mathematically as a polynomial. Namely, if $x$ is a plaintext, then ciphertext $y$ is equal to $p(x)$, where $p$ is a polynomial with integer coefficients.

Bob's employer decided to test it. At first, he encrypted the number 20 and obtained the number~7. Secondly, he encrypted the number 15 and obtained the number 5. After that he said to Bob that there was a mistake in the implementation of the algorithm and did not hire him. What was wrong?

\subsubsection{Solution}

Let $p(x) = c_0 + c_1 x + \ldots + c_n x^n$. Then $p(a) - p(b) = c_1 (a - b) + \ldots + c_n (a^n - b^n)$, where $a,b$ are some integers. Since $(a^k-b^k)$ is divided by $(a-b)$, we have that $p(a) - p(b)$ is divided by $(a-b)$.
By condition, we have $p(20) = 7$ and $p(15) = 5$, but $5$ does not divide $2$. Hence, there is a mistake in the implementation.
Almost all the participants solved the problem.

\subsection{Problem ``A secret house''}
\hypertarget{pr-house}{}
\subsubsection{Formulation}

\begin{figure}[!h]
	\centering
	\begin{tabular}{cc}
		\includegraphics[width=0.15\textwidth]{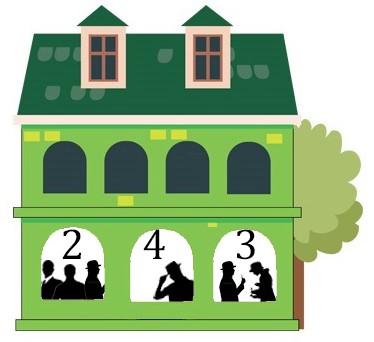}
		&
		\includegraphics[width=0.6\textwidth]{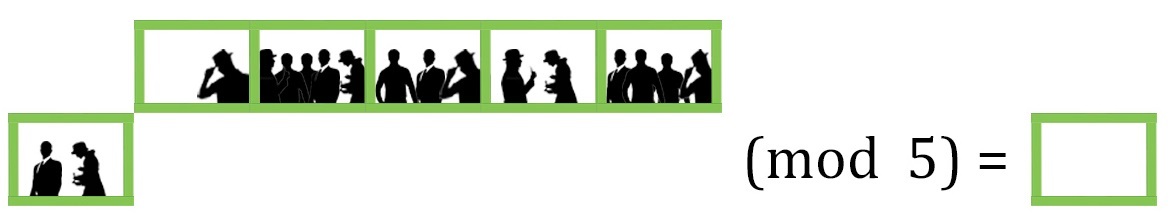}
		\\
		(a) & (b)
	\end{tabular}
	\caption{{\small A secret house}}
	\label{fig:house}
\end{figure}

You can see a secret house in Fig.~\ref{fig:house}(a).
Looking on it, could you understand
what should be shown inside the frame left blank in Fig.~\ref{fig:house}(b)?

\subsubsection{Solution}

Looking on the house, one can see that the number in a window is equal to ``5 minus the number of shadows'' inside the window. Hence, we can guess that the task is to calculate $3^{40231} \pmod 5$.
Since $3^4 = 1 \pmod 5$, then $3^{40231}\pmod 5 = 3^{4\cdot 10057 + 3} \pmod 5 = 3^3 \mod 5 = 4$. Hence, there should be one shadow inside the frame.

\subsection{Problem ``RGB''}
\hypertarget{pr-RGB}{}
\subsubsection{Formulation}

Victor is studying the Moctod search server. Inside its software,
he found two integer variables $a$ and $b$ that change their values when
special search queries ``\texttt{RED}'', ``\texttt{GREEN}'' and ``\texttt{BLUE}'' are processed.
More precisely, the pair $(a,b)$ is changed to $(a+18b,18a-b)$
when processing the query ``\texttt{RED}'',
to $(17a+6b,-6a+17b)$ when processing ``\texttt{GREEN}'',
and to $(-10a-15b,15a-10b)$ when processing ``\texttt{BLUE}''.
When any of $a$ or $b$ reaches a multiple of $324$, it resets to $0$.
Whenever $(a,b)=(0,0)$, the server crashes.

On the server startup, the variables $(a,b)$ are set to $(20,20)$. Prove that the
server will never crash with these initial values, regardless of the search
queries processed.

\subsubsection{Solution}

The number $ 325 $ is the first natural number that can be written as
sums of squares in three different ways (up to permutation of terms):
$$
325 = 1^2 + 18^2 = 6^2 + 17^2 = 10^2 + 15^2.
$$
Keeping this in mind, if $(A, B)$ is the result of changing $(a, b)$ with some query, then
$$
A^2 + B^2 = 325 (a^2 + b^2) \equiv a^2 + b^2 \pmod {324}.
$$
Thus, the number $(a^2 + b^2) \bmod 324 $ does not change for any chain of queries
(in order words, it is an invariant).
Since initially $(20^2 + 20^2) \bmod 324 = 152 \neq 0 $, the server will never crash.

\subsection{Problem ``Miller~---~Rabin revisited''}\label{miller-rabin}
\hypertarget{pr-miller-rabin}{}
\subsubsection{Formulation}

Bob decided to improve the famous Miller~---~Rabin primality test and invented his test given in Algorithm~\ref{alg:MRm}.
The odd number~$n$ being tested is represented in the form $n - 1 =2^k 3^{\ell} m $,
where $m$ is not divisible by~$2$ or~$3$.

\begin{algorithm}
	\caption{Bob's primality test}\label{alg:MRm}
	\begin{enumerate}[noitemsep]
		\item[{\bf 1.}]
		Take a random $a\in\{2,\ldots,n-2\}$.
		\item[{\bf 2.}]
		Put $a\leftarrow a^m \bmod n$.
		If $a=1$, return ``{\tt PROBABLY PRIME}''.
		\item[{\bf 3.}]
		For $i=0,1,\ldots,{\ell}-1$ do the following steps:
		\begin{enumerate}[noitemsep]
			\item
			$b\leftarrow a^2 \bmod n$;
			\item
			if $a+b+1$ is divisible by~$n$, return ``{\tt PROBABLY PRIME}'';
			\item
			$a\leftarrow ab \bmod n$.
		\end{enumerate}
		\item[{\bf 4.}]
		For $i=0,1,\ldots,k-1$ repeat:
		\begin{enumerate}[noitemsep]
			\item
			if $a+1$ is divisible by~$n$, return ``{\tt PROBABLY PRIME}'';
			\item
			$a\leftarrow a^2 \bmod n$.
		\end{enumerate}
		\item[{\bf 5.}]
		Return ``{\tt COMPOSITE}''.
	\end{enumerate}
\end{algorithm}

\begin{enumerate}
	\item[{\bf Q1}] {
		Prove that Algorithm~\ref{alg:MRm} does not fail, that is, not return ``{\tt COMPOSITE}'',
		for a prime~$n$.
	}
	\item[{\bf Q2}]
	\underline{\bf Bonus problem (extra scores, a special prize!)}
	
	A composite integer $n$ may be classified as ``{\tt PROBABLY PRIME}'' by a mistake.
	It is known that for the usual Miller --- Rabin test the error probability is
	less than~$1/4$. Can this estimation be improved when we are switching to Algorithm~\ref{alg:MRm}?
\end{enumerate}

\noindent{\bf Remark.} The expression $a\leftarrow a^m \bmod n$ means that $a$ takes a new value that is equal to the remainder of dividing $a^m$ by $n$.

\subsubsection{Solution}

Let us prove that Algorithm~\ref{alg:MRm} does not fail ({\bf Q1}).

If $n$ is prime, then by Fermat's Little Theorem $n$ divides
\begin{align*}
	a^{n-1}-1 &= a^{2^k 3^l m} - 1 = (a^{2^{k-1} 3^l m} - 1)(a^{2^{k-1} 3^l m} + 1)=\ldots=\\
	&=(a^{3^l m} - 1)\prod_{i=0}^{k-1}\left(a^{2^i 3^l m} + 1\right)=
	((a^{3^{l-1} m})^3 - 1)\prod_{i=0}^{k-1}\left(a^{2^i 3^l m} + 1\right)=\\
	&=(a^{3^{l-1} m} - 1)((a^{3^{l-1} m})^2 + a^{3^{l-1} m} + 1)
	\prod_{i=0}^{k-1}\left(a^{2^i 3^l m} + 1\right)=\ldots=\\
	&=(a^m-1)\prod_{j=0}^{l-1}\left((a^{3^j m})^2 + a^{3^j m} + 1\right)
	\prod_{i=0}^{k-1}\left(a^{2^i 3^l m} + 1\right).
\end{align*}
A prime number $n$ must divide one of the parentheses in the last expression.
The required statement follows from this.

The answer for the question {\bf Q2} is ``the estimation is not improved''. Let us prove this.
In the original Miller --- Rabin test, instead of steps 2 and 3, the following step is performed:
\begin{enumerate}
	\item[{\bf 23.}]
	$ a\leftarrow a^{3^lm} \bmod n $.
	If $ a = 1 $, return ``{\tt PROBABLY PRIME}''.
\end{enumerate}

In other words, the following congruence relation is checked:
\begin{equation} \label{Eq.MR3.1}
	a^{3^lm} \equiv 1 \pmod{n}.
\end{equation}
If (\ref{Eq.MR3.1}) is satisfied, then $A = a^{3^{l-1}m}$
is the cube root of $1$ modulo~$n$:
$$
A^3-1\equiv 0\pmod{n}\quad\Leftrightarrow\quad
(A-1)(A^2+A+1)\equiv 0\pmod{n}.
$$
In this case, either $ A \equiv 1 \pmod{n} $, i.e. ~
\begin{equation}\label{Eq.MR3.2}
	a^{3^{l-1} m}\equiv 1\pmod{n},
\end{equation}
or
$A^2+A\equiv -1\pmod{n}$.
Both cases are analyzed in Bob's test.
In the first case, the congruence relation~\eqref{Eq.MR3.2}
is analyzed in the same way as \eqref{Eq.MR3.1}.

Thus, the answer ``{\tt PROBABLY PRIME}'' in Miller --- Rabin test
is returned if and only if the same answer is returned in
Bob's test.
Bob's test has an advantage over Miller~---~Rabin test. It is more efficient since the correctness of \eqref{Eq.MR3.1} can be obtained earlier.

The question {\bf Q2} was correctly solved by 10 participants and teams. They are Artur Puzio (Poland), Leo Boitel (France), Geng Wang (China), Gabor~P.~Nagy (Hungary),  the team of Albert Smith, Ethan Tan, Guowen Zhang (Australia), the team of Mircea-Costin Preoteasa, Gabriel Tulba-Lecu, Ioan Dragomir (Romania), the team of Sergey Bystrevskii, Maksim Starodubov, Evgeny Mikhalchuk (Russia), the team of Mohammad Akbarizadeh, Reza Kaboli, Sajjad Bagheri (Iran), the team of Jeremy Jean, Hugues Randriam (France), Irina Slonkina (Russia).

\subsection{Problem ``Mysterious event''}
\hypertarget{pr-event}{}
\subsubsection{Formulation}

Mr. Bob is the editor in-chief of a well known magazine. He has many interests and activities in addition to work: meetings with bright people of politics and art, dancing, fishing, and even stenography and linguistics.

Every week, the magazine publishes a hard Sudoku on the last page. Mr. Bob likes this game too! So, it is a pleasure for him to personally analyze all solutions from the readers. He sits down in his office with a cup of coffee and looks through all the PNG-files with photos of solutions.

\begin{figure}[!h]
	\centering
	\includegraphics[width=0.33\textwidth]{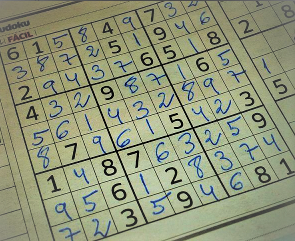}
	\vspace{-0.2cm}
	\caption{{\small Sudoku}}
	\label{fig:sudoku}
\end{figure}

But suddenly Mr. Bob disappeared. The last solution he could see on his monitor was that in Fig.~\ref{fig:sudoku} (\href{http://nsucrypto.nsu.ru/media/MediaFile/Mysterious_event-email.png}{\textcolor{blue}{here}}
is a {\bf link} to it, if you are interested in).

But what happened? Where is Mr. Bob?

\subsubsection{Solution}

As Mr. Bob likes stenography and the format of the given file is png, one can try to find message hidden in Fig.~\ref{fig:sudoku} using steganography tools, for example \cite{stego}. It reveals the message ``They know that you are a spy! Get back to the center right now.''
So, Mr.Bob is in the center.

\subsection{Problem ``CPA game''}
\hypertarget{pr-cpa}{}
\subsubsection{Formulation}

Suppose we have a system for the encryption of binary messages. The system has the following characteristics:
\begin{itemize}[noitemsep]
	\item
	Every message is divided into blocks of length $n$ that are called plaintexts (it is supposed that the length of messages is divisible by $n$).
	\item
	The system employs a block cipher with the encryption function $E$ in cipher block chaining (CBC) mode (see the picture below). A block, an initialization vector $IV$ and a key lengths are equal to $n$. The result of encryption of the message is a concatenation of $IV$ and the ciphertexts of all plaintexts it consists of.
	\item
	The $IV$ for the first message is chosen randomly by using a secure pseudo\-random number generator. The last ciphertext block of the $i$-th message is used as the $IV$ for the $(i+1)$-st message.
\end{itemize}

Let Alice be an honest user of the system. Victor, an adversary, convinced her to play {\bf chosen--plaintext attack game} (CPA game) with him.

\bigskip
\noindent The game is the following:
\begin{enumerate}[noitemsep]
	\item[{\bf 1.}]
	Alice selects a key $k \in \{0,1\}^{n}$ and chooses a bit $b \in \{0,1\}$.
	\item[{\bf 2.}] Victor submits a sequence of $q$ queries to Alice.
	For $i = 1, 2,\ldots, q$ repeat
	\begin{enumerate}[noitemsep]
		\item Victor chooses a pair of messages, $m_{i,0}, m_{i,1}$ of the same length.
		\item Alice encrypts $m_{i,b}$ with the key $k$ and gets $c_i$ (that is the sequence of corresponding $IV$ and ciphertexts). She sends $c_i$ to Victor.
	\end{enumerate}
	\item[{\bf 3.}]
	Victor outputs a bit $b^{*} \in \{0, 1\}$.
\end{enumerate}

Let W be the event that Victor guesses the bit, that is $b^{*} = b$. We define Victors's advantage
with respect to $E$ as ${\rm CPAadv}:= |{\rm Pr}[{\rm W}] - 1/2|$.
Victor wins the game if he can build an efficient algorithm such that CPAadv is not negligible.

\bigskip

\noindent {\bf Task}.
Construct an efficient probabilistic polynomial-time (PPT) algorithm that wins the CPA game against this implementation with an advantage close to 1/2.

\subsubsection{Solution}

We describe two deterministic algorithms that win the given CPA game with two queries in Algorithms~\ref{alg:CPA1},\ref{alg:CPA2}. Let ${\bf 0}$ and ${\bf 1}$ denote all zeros and all ones vectors from the space $\mathbb{F}_2^n$.

\begin{algorithm}
	\caption{The first deterministic algorithm}\label{alg:CPA1}
	\begin{itemize}[noitemsep]
		\item[\textbf{q1:}]
		\begin{itemize}[noitemsep]
			\item[(a)] Victor chooses a pair of messages $m_{1,0}=m_{1,1}={\bf 0}$ and sends them to Alice;
			\item[(b)] Alice sends $c_1=\left(IV,E_k(IV)\right)$ to Victor;
		\end{itemize}
		\item[\textbf{q2:}]
		\begin{itemize}[noitemsep]
			\item[(a)] Victor chooses a pair of messages	
			$m_{2,0} =IV\oplus E_k(IV)$, $m_{2,1} =IV\oplus E_k(IV)\oplus{\bf 1}$
			and sends them to Alice;
			\item[(b)] Alice sends $c_2=\left(E_k(IV),C\right)$ to Victor. Depending on the value of~$b$, the ciphertext~$C$ is equal to~$E_k(IV)$ if~$b=0$, and it holds $C=E_k(IV\oplus{\bf 1})$ if~$b=1$.
		\end{itemize}
	\end{itemize}
	Finally, Victor outputs $b^*=0$ if $C=E_k(IV)$ and $b^*=1$ otherwise.
\end{algorithm}

\begin{algorithm}
	\caption{The second deterministic algorithm}\label{alg:CPA2}
	\begin{itemize}[noitemsep]
		\item[\textbf{q1:}]
		\begin{itemize}[noitemsep]
			\item[(a)] Victor chooses a pair of messages $m_{1,0}={\bf 0}$, $m_{1,1}={\bf 1}$ and sends them to Alice;
			\item[(b)] Alice sends $c_1=\left(IV,C\right)$ to Victor, where the ciphertext~$C$ is equal to~$E_k(IV)$ if~$b=0$, and it holds $C=E_k(IV\oplus{\bf 1})$ if~$b=1$;
		\end{itemize}
		\item[\textbf{q2:}]
		\begin{itemize}[noitemsep]
			\item[(a)] Victor chooses a pair of messages $m_{2,0}=m_{2,1}=IV\oplus C$
			and sends them to Alice;
			\item[(b)] Alice sends $c_2=\left(E_k(IV),E_k(IV)\right)$ to Victor.
		\end{itemize}
	\end{itemize}
	Finally, Victor outputs $b^*=0$ if $C=E_k(IV)$ and $b^*=1$ otherwise.
\end{algorithm}

There were several solutions from the participants that proposed the approaches described above, as well as many 3-queries deterministic and probabilistic algorithms.

\subsection{Problem ``Stairs-Box''}
\hypertarget{pr-stairs}{}
\subsubsection{Formulation}

Nicole was climbing stairs and has found a box containing a curious permutation on the set of elements $\{{\tt 0}, {\tt 1},\ldots, {\tt 63}\}$:
\medskip

\indent\indent\indent\indent\indent $S=$ \texttt{[}\,\,\,
\texttt{13},\texttt{18},\texttt{20},\texttt{55},\texttt{23},\texttt{24},\texttt{34},\texttt{ 1},\texttt{62},\texttt{49},\texttt{11},\texttt{40},\texttt{36},\texttt{59},\texttt{61},\texttt{30},\\
\indent\indent\indent\indent\indent\phantom{xxxx}\ \ \ \texttt{33},\texttt{46},\texttt{56},\texttt{27},\texttt{41},\texttt{52},\texttt{14},\texttt{45},\texttt{ 0},\texttt{29},\texttt{39},\texttt{ 4},\texttt{ 8},\texttt{ 7},\texttt{17},\texttt{50},\\
\indent\indent\indent\indent\indent\phantom{xxxx}\ \ \
\texttt{ 2},\texttt{54},\texttt{12},\texttt{47},\texttt{35},\texttt{44},\texttt{58},\texttt{25},\texttt{10},\texttt{ 5},\texttt{19},\texttt{48},\texttt{43},\texttt{31},\texttt{37},\texttt{ 6},\\
\indent\indent\indent\indent\indent\phantom{xxxx}\ \ \ \texttt{21},\texttt{26},\texttt{32},\texttt{ 3},\texttt{15},\texttt{16},\texttt{22},\texttt{53},\texttt{38},\texttt{57},\texttt{63},\texttt{28},\texttt{60},\texttt{51},\texttt{ 9},\texttt{42} \,\,\, \texttt{]}
\medskip

\noindent So, the element {\tt 0} it maps to {\tt 13}, the element {\tt 1} to {\tt 18}, etc.

Nicole understands that it is possible to consider such a permutation as a vectorial Boolean function $S:\mathbb{F}_2^6 \to \mathbb{F}_2^6$ if every number between {\tt 0} and {\tt 63} one replaces with a binary vector of length~6. For instance, $S(000010) = (010100)$, since $S$ maps $2$ to~$20$. She knows that $S$ can be given in terms of coordinate functions as $S(x) = (s_1(x),\ldots,s_6(x))$, and each Boolean function $s_i$ can be represented in the algebraic normal form using binary operations XOR and AND in the following way: $s_i(x) = \bigoplus_{I\in \mathcal{P}(N)} a_{I} \big(\prod_{i\in I} x_i \big),$
where $\mathcal{P}(N)$ is the power set of $N=\{1,\ldots,6\}$ and $a_I\in\mathbb{F}_2$.

A label on the box said that the function $S$ can be represented as a composition of three maps in the following way:
$$
S = A \circ X \circ B,
$$
where $A,B:\mathbb{F}_2^6 \to \mathbb{F}_2^6$ are {\bf linear maps} and $X$ is a function with a {\bf short arithmetic expression modulo $64$}. Nicole knows that a linear map over $\mathbb{F}_2^6$ can be defined by multiplication with a $6\times6$ matrix over $\mathbb{F}_2$. But she wonders what is supposed by ``a~short arithmetic expression modulo 64''? Probably, Nicole also should consider maps as classical modular operations such as addition, substraction, multiplication modulo 64?..

Help Nicole to find the secret function $X$ and the respective maps $A,B$!

\subsubsection{Solution}

Arithmetic operations modulo $2^6$ can be reduced modulo smaller powers of 2. Most importantly, the output modulo 2 depends only on the input modulo 2 (1 bit), the output modulo $2^i$ depends only on the input modulo $2^i$ ($i$ input bits, $1 \leqslant i \leqslant 6$).

It follows that there must exist linear combinations of outputs of $S$ with algebraic degrees less or equal to each of 1, 2, 3, 4, 5, 5 (``staircase''). And indeed, such combinations do exist for the given S-box $S$.
While there is some freedom left in choosing such combinations, the number of possibilities is reasonably small. Any such choice identifies a candidate for the linear map $A$. The same idea can be applied to $S^{-1}$ to obtain candidates for $B$. Using the fact that $i$ least significant bits of the output of $X$ must depend only on $i$ least significant bits of the input of $X$, correct candidates for $A, B$ can be recovered in a sequential bit-by-bit manner.

There exist 8 solutions, any of which was accepted as a correct answer:
\begin{align*}
	X: \mathbb{Z}_{64} \to \mathbb{Z}_{64},~X(x) \in \{
	& x+1,\ x+17,\ x+33,\ x+49,\\
	& 33x+1,\ 33x+17,\ 33x+33,\ 33x+49
	\}.
\end{align*}
In total, 15 teams managed to solve this problem completely and 12 teams got only partial progress. Many teams guessed the linear shape of the polynomial of $X$ and used creative ways to verify their guess. Teams of Gongyu Shi, Xinzhou Wang, Yu-hang Jii	(China) and Weidan Ji, Wenwen Xia, Zhang Hongyi (China) used the Walsh spectrum exploiting its invariance under composition of the function with linear maps and further recovered $A,B$ efficiently by matching the rows/columns of the Linear Approximation Tables (LAT) of $S$ and $X$. The team of Gyumin Roh, Hyunsik Jeong, Mincheol Son (South Korea) developed similar method but using Difference Distribution Table (DDT) instead of the LAT. Hieu Nguyen Duy (Vietnam) used more direct approach to reconstructing $A,B$ row-by-row/column-by-column with the constraint of the partial solution $X$ modulo $2^i$ having the form linear polynomial $x\mapsto ax+b$.

\subsection{Problem ``Hidden RSA''}
\hypertarget{pr-RSA}{}
\subsubsection{Formulation}

Bob has learned about the public-key cryptography and now anyone can send a
secret message to him. The message is encoded by a nonnegative integer $x$ which has
at most $70$ digits in the decimal representation. To send a message for Bob,
one has to enter it on his  \href{https://nsucrypto.nsu.ru/archive/2020/round/2/task/3/}{\textcolor{blue}{webpage}} \cite{nsucrypto-rsa}. After the message is entered,
it is immediately encrypted using RSA. The encryption result is
$$
\texttt{Encr}(x)=x^e \bmod n,
$$
where $n$ is a modulus (product of two distinct odd primes~$p$ and~$q$)
and $e$ is a public exponent (coprime with $p-1$ and $q-1$).
Bob is afraid of hackers and does not disclose either~$n$ or~$e$ (even though
this contradicts the usual usage of the RSA cryptosystem).

Victor has intercepted the encrypted message
\begin{equation*}
	\resizebox{\textwidth}{!}{
		$y=71511896681324833458361392885184344933333159830863878600189212073777582178173,$
	}
\end{equation*}
which Alice has sent to Bob.

Help Victor to decrypt~$y$. You can enter any allowed message~$x$ on the Bob's
\href{https://nsucrypto.nsu.ru/archive/2020/round/2/task/3/}{\textcolor{blue}{website}} \cite{nsucrypto-rsa} and receive in response the corresponding ciphertext~$\texttt{Encr}(x)$.

\subsubsection{Solution}

Victor takes advantage of the fact that RSA typically uses a small open exponent~$e$. Victor views small candidate exponents $\hat{e} = 3,5, \ldots $, searching for the correct one among them and at the same time determining~$n$.

Viktor processes $ \hat{e}$ as follows.
First, he checks the condition $ 2^{\hat{e}} \geqslant \texttt{Encr}(2) $.
If the condition is not satisfied, then $ \hat{e}$ is rejected.
Second, Victor defines $ \hat{n} = 2^e-\texttt{Encr}(2)$.
This is an estimate of the modulus $n$ in the sense that if $ \hat{e} = e $, then
$ \hat{n}$ is a multiple of~$n$.
Third, for several random $x$ Victor refines the estimate:
$$
\hat{n}\leftarrow\gcd(\hat{n},(x^{\hat{e}}\bmod \hat{n})-\texttt{Encr}(x)).
$$
If $ \hat{e} = e $, then the estimate $ \hat{n}$ quickly converges to~$n$.
If $ \hat{e} \neq e $, then $ \hat{n} $ quickly converges to~$1$.

Using the method described above, Victor finds $e = 65537$ and
\begin{align*}
n &= 76200708443433250012501342992033571586971760218934756930058661627867825188509.
\end{align*}
The module $n$ ($256$-bit) can be quickly factorized using programs like
\texttt{msieve} or \texttt{cado-nfs}.

As a result, prime divisors can be found
\begin{align*}
	p & = 232086664036792751646261018215123451301, \\
	q & = 328328681700354546732404725320581286809.
\end{align*}
Then the secret exponent is determined
\begin{align*}
	d & = e ^ {- 1} \bmod (p-1) (q-1) = \\
	& =  58041460011714671214337771652949080061981291861469879231637604933853779098273
\end{align*}
and the desired message
$$
y^d \bmod n = 202010181600.
$$
This is the NSUCRYPTO'2020 start time code (October 18, 2020, 16:00).

\subsection{Problem ``Orthomorphisms''}
\hypertarget{pr-ortho}{}
\subsubsection{Formulation}

A young cryptographer Bob wants to build a new block cipher based on the Lai-Massey scheme. The Lai-Massey scheme depends on a finite group $G$ with the neutral element $e$ and an orthomorphism of $G$. Bob decides to use a nonabelian group and chooses a dihedral group ${D_{{2^m}}}$, $m \geqslant 4$, generated by $a,u$ with presentation
$${a^{{2^{m - 1}}}} = e,\ {u^2} = e,\ ua = {a^{ - 1}}u.$$

Let $\theta $ be a permutation of a finite group $G$. Then $\theta $ is called an {\bf orthomorphism of $G$} if the mapping $\pi :\alpha  \mapsto {\alpha ^{ - 1}}\theta (\alpha )$ is a permutation of $G$.

Bob needs to construct an orthomorphism of ${D_{{2^m}}}$. He considers the set ${\rm{DM}}_m$ consisting of all mappings $\theta _{({q_1},{q_2},{b_1},{b_2})}^{({r_1},{r_2},{c_1},{c_2})}$ on ${D_{{2^m}}}$ given by
\begin{align*}\theta _{(q_1,q_2,b_1,b_2)}^{(r_1,r_2,c_1,c_2)}:{a^i} & \mapsto
\begin{cases}
	a^{{r_1}i + c_1}  &\text{if $i \in \{ {0,\ldots,2^{m - 2}} - 1\} $,} \\
	{a^{{r_2}i + {c_2}}}u &\text{if $i \in \{ {2^{m - 2}}{,\ldots,2^{m - 1}} - 1\}$,}  \\
\end{cases}
\\
\theta _{(q_1,q_2,b_1,b_2)}^{(r_1,r_2,c_1,c_2)}:{a^i}u & \mapsto
\begin{cases}
	{a^{{q_1}i + {b_1}}}u, &\text{if $i \in \{ {0,\ldots,2^{m - 2}} - 1\} $,} \\
	a^{{q_2}i + {b_2}}, &\text{if $i \in \{ {2^{m - 2}}{,\ldots,2^{m - 1}} - 1\}$,}  \\
\end{cases}
\end{align*}
and depending on ${b_i},{c_i},{r_i},{q_i} \in \{ {0,\ldots,2^{m - 1}} - 1\} $ for $i \in \{ 1,2\} $, where the operations addition and multiplication are over the residue ring $\mathbb{Z}_{{2^{m - 1}}}$.

\begin{itemize}
	\item[{\bf Q1}] Let $m = 4$. Help Bob to describe all orthomorphisms of ${\rm{D}}{{\rm{M}}_m}$ and find their number.
	
	\item[{\bf Q2}] For each $m \geqslant 4$, help Bob to describe all orthomorphisms of ${\rm{D}}{{\rm{M}}_m}$, i.\,e. give necessary and sufficient conditions on ${b_i},{c_i},{r_i},{q_i}$ for $i \in \{ 1,2\} $ such that $\theta _{({q_1},{q_2},{b_1},{b_2})}^{({r_1},{r_2},{c_1},{c_2})}$ is an orthomorphism of~${D_{{2^m}}}$.
\end{itemize}

\subsubsection{Solution}

Let ${Z_n} = \{ 0,...,n - 1\} $ for a positive integer $n\geqslant 1$.

\smallskip
\noindent{\bf Theorem.} Let $m\geqslant 4$. A mapping $\theta _{({q_1},{q_2},{b_1},{b_2})}^{({r_1},{r_2},{c_1},{c_2})} \in {{\rm{D M }}_m}$ is an orthomorphism if and only if  ${b_i},{c_i},{r_i},{q_i} \in {Z_{{2^{m - 1}}}}$ for $i \in \{ 1,2\} $ satisfy one of the following conditions:
%
%
%

\begin{enumerate}[noitemsep]
  \item {If ${r_1} \equiv {r_2} \equiv 3\;(\bmod \;4)$, then
$r_1 = q_2, r_2 = q_1, \; c_1 = b_2, c_2 = b_1, \; c_1 + c_2\; \equiv 1\;(\bmod \;2).$
  }
  \item {If ${r_1} \equiv {r_2} \equiv 2\;(\bmod \;4)$, then
${r_1} = {q_1}, \; {r_2} = {q_2}, \;$

\hspace{4.9cm}$ {q_1} - 1 \equiv {b_1} + {c_1}\;(\bmod \;{2^{m - 1}}), \;
 {q_2} - 1 \equiv {b_2} + {c_2}\;(\bmod \;{2^{m - 1}}),$

\hspace{4.8cm}
${b_1} + {c_2}\; \equiv 1\;(\bmod \;2), \; {b_2} + {c_1}\; \equiv 1\;(\bmod \;2).$
  }
\end{enumerate}

\noindent{\bf Proof of Theorem}.
Let $\theta  = \theta _{({q_1},{q_2},{b_1},{b_2})}^{({r_1},{r_2},{c_1},{c_2})}$. It is clear that $\theta $ is a permutation if and only if

$$\bigcup\limits_{j = 0}^{{2^{m - 2}} - 1} {\left\{ {{r_1}j + {c_1}\;} \right\}}  \cap \bigcup\limits_{j = {2^{m - 2}}}^{{2^{m - 1}} - 1} {\left\{ {{q_2}j + {b_2}} \right\}}  = \emptyset, ~~~~~
\bigcup\limits_{j = 0}^{{2^{m - 2}} - 1} {\left\{ {{r_1}j + {c_1}\;} \right\}}  \cup \bigcup\limits_{j = {2^{m - 2}}}^{{2^{m - 1}} - 1} {\left\{ {{q_2}j + {b_2}} \right\}}  = {Z_{{2^{m - 1}}}},$$
$$\bigcup\limits_{j = {2^{m - 2}}}^{{2^{m - 1}} - 1} {\left\{ {{r_2}j + {c_2}} \right\}}  \cap \bigcup\limits_{j = 0}^{{2^{m - 2}} - 1} {\left\{ {{q_1}j + {b_1}} \right\}}  = \emptyset, ~~~~~
\bigcup\limits_{j = {2^{m - 2}}}^{{2^{m - 1}} - 1} {\left\{ {{r_2}j + {c_2}} \right\}}  \cup \bigcup\limits_{j = 0}^{{2^{m - 2}} - 1} {\left\{ {{q_1}j + {b_1}} \right\}}  = {Z_{{2^{m - 1}}}},$$
%
where the operations addition and multiplication are over the residue ring $\mathbb{Z}_{{2^{m - 1}}}$.
They are equivalent to conditions
\begin{subequations}\label{Eqorth1}
	\begin{align}
		{r_1}{j_1} - {q_2}{j_2} &\not  \equiv {q_2}{2^{m - 2}} + {b_2} - {c_1}\;(\bmod \;{2^{m - 1}}), \label{E:5.12}\\
		{r_2}{j_1} - {q_1}{j_2} &\not  \equiv {q_1}{2^{m - 2}} + {b_1} - {c_2}\;(\bmod \;{2^{m - 1}}), \label{E:5.13}\\
		{r_1}({j'_1} - {j'_2}) &\not  \equiv 0\;(\bmod \;{2^{m - 1}}),
		\label{E:5.14}\\
		{r_2}({j'_1} - {j'_2}) &\not  \equiv 0\;(\bmod \;{2^{m - 1}}),
		\label{E:5.15}\\
		{q_1}({j'_1} - {j'_2}) &\not  \equiv 0\;(\bmod \;{2^{m - 1}}),
		\label{E:5.16}\\
		{q_2}({j'_1} - {j'_2}) &\not  \equiv 0\;(\bmod \;{2^{m - 1}}),
		\label{E:5.17}
	\end{align}
\end{subequations}
which hold for all ${j_1},{j_2} \in {Z_{{2^{m - 2}}}}$ and all ${j'_1},{j'_2} \in {Z_{{2^{m - 2}}}}$ with ${j'_1} \ne {j'_2}$.

From conditions (\ref{E:5.14}) -- (\ref{E:5.17}), it follows that
\begin{equation}\label{E:5.18}
	{r_1}\not  \equiv 0\;(\bmod \;4), \; {r_2}\not  \equiv 0\;(\bmod \;4), \; {q_1}\not  \equiv 0\;(\bmod \;4), \; {q_2}\not  \equiv 0\;(\bmod \;4).
\end{equation}

Note that  $\pi :\alpha  \mapsto {\alpha ^{ - 1}}\theta (\alpha )$ is given by
\begin{align*}
\pi :{a^i} \mapsto
& \begin{cases}
	a^{({r_1} - 1)i + {c_1}}     &\text{if $i \in {Z_{{2^{m - 2}}}} $,} \\
	{a^{({r_2} - 1)i + {c_2}}}u  &\text{if $i \in \{ {2^{m - 2}}{,...,2^{m - 1}} - 1\} $,}
\end{cases}
\\
\pi :{a^i}u \mapsto
&\begin{cases}
	a^{- ({q_1} - 1)i - {b_1}}  &\text{if $i \in {Z_{{2^{m - 2}}}} $,} \\
	{a^{ - ({q_2} - 1)i - {b_2}}}u  &\text{if $i \in \{ {2^{m - 2}}{,...,2^{m - 1}} - 1\} $,}
\end{cases}
\end{align*}
where the operations addition, multiplication and subtraction are over ${\mathbb{Z}_{{2^{m - 1}}}}$.

For each $i \in \{ 1,2\}, $ we suppose
${\tilde r_i} = {r_i} - 1\,\bmod \,{2^{m - 1}}, \; {\tilde q_i} = 1 - {q_i}\,\bmod \,{2^{m - 1}}, \; {\tilde b_i} = {2^{m - 1}} - {b_i}.$

It is clear that $\pi $ is a permutation if and only if
\newpage
$$\bigcup\limits_{j = 0}^{{2^{m - 2}} - 1} {\left\{ {{{\tilde r}_1}j + {c_1}} \right\}}  \cap \bigcup\limits_{j = 0}^{{2^{m - 2}} - 1} {\left\{ {{{\tilde q}_1}j + {{\tilde b}_1}} \right\}}  = \emptyset, ~~~~~
\bigcup\limits_{j = 0}^{{2^{m - 2}} - 1} {\left\{ {{{\tilde r}_1}j + {c_1}} \right\}}  \cup \bigcup\limits_{j = 0}^{{2^{m - 2}} - 1} {\left\{ {{{\tilde q}_1}j + {{\tilde b}_1}} \right\}}  = {Z_{{2^{m - 1}}}},$$
$$\bigcup\limits_{j = {2^{m - 2}}}^{{2^{m - 1}} - 1} {\left\{ {{{\tilde r}_2}j + {c_2}} \right\}}  \cap \bigcup\limits_{j = {2^{m - 2}}}^{{2^{m - 1}} - 1} {\left\{ {{{\tilde q}_2}j + {{\tilde b}_2}} \right\}}  = \emptyset, ~~~~~
\bigcup\limits_{j = {2^{m - 2}}}^{{2^{m - 1}} - 1} {\left\{ {{{\tilde r}_2}j + {c_2}} \right\}}  \cup \bigcup\limits_{j = {2^{m - 2}}}^{{2^{m - 1}} - 1} {\left\{ {{{\tilde q}_2}j + {{\tilde b}_2}} \right\}}  = {Z_{{2^{m - 1}}}}, $$
where the operations addition and multiplication are over the residue ring $\mathbb{Z}_{{2^{m - 1}}}$.

They are equivalent to conditions
\begin{subequations}\label{Eqorth2}
	\begin{align}
		({r_1} - 1){j_1} - (1 - {q_1}){j_2} &\not  \equiv  - {b_1} - {c_1}\;(\bmod \;{2^{m - 1}}),
		\label{E:5.30}\\
		({r_2} - 1){j_1} - (1 - {q_2}){j_2} &\not  \equiv  - {b_2} - {c_2}\;(\bmod \;{2^{m - 1}}),
		\label{E:5.31}\\
		({r_1} - 1)({j'_1} - {j'_2}) &\not  \equiv 0\;(\bmod \;{2^{m - 1}}),
		\label{E:5.32}\\
		({r_2} - 1)({j'_1} - {j'_2}) &\not  \equiv 0\;(\bmod \;{2^{m - 1}}),
		\label{E:5.33}\\
		({q_1} - 1)({j'_1} - {j'_2}) &\not  \equiv 0\;(\bmod \;{2^{m - 1}}),
		\label{E:5.34}\\
		({q_2} - 1)({j'_1} - {j'_2}) &\not  \equiv 0\;(\bmod \;{2^{m - 1}}),
		\label{E:5.35}
	\end{align}
\end{subequations}
which hold for all ${j_1},{j_2} \in {Z_{{2^{m - 2}}}}$ and all $j'_1,j'_2 \in {Z_{{2^{m - 2}}}}$ with ${j'_1} \ne {j'_2}$.

From conditions (\ref{E:5.32}) -- (\ref{E:5.35}), it follows that
\begin{equation}\label{E:5.36}
	{r_1}\not  \equiv 1\;(\bmod \;4), \; {r_2}\not  \equiv 1\;(\bmod \;4), \; {q_1}\not  \equiv 1\;(\bmod \;4), \; {q_2}\not  \equiv 1\;(\bmod \;4).
\end{equation}

Then we will use the following Lemma.

\medskip
\noindent\textbf{Lemma}. Let $d \geqslant 4$, ${R^{(d)}} = \left\{ {r \in {Z_{{2^{d - 1}}}}|r \equiv t\;(\bmod \;4),\;t \in \{ 1,2,3\} } \right\},$ and\\
${\bar A^{(d)}}({h_1},{h_2}) = \left\{ {{h_1}{j_1} - {h_2}{j_2}\; \bmod{2^d} \;|{j_1},{j_2} \in {Z_{{2^{d - 1}}}}} \right\},\; {h_1},{h_2} \in {R^{(d)}}. $

Then
$$
{\bar A^{(d)}}({h_1},{h_2}) =
\begin{cases}
	Z_{{2^d}}\backslash \{ {2^{d - 1}}\} &\text{if ${h_1} = {h_2},\;\;{h_1} \equiv {h_2} \equiv 1\;(\bmod \;2)$,} \\
	{Z_{2^d}}\backslash \{ {h_2}\}      &\text{if ${h_2} = {2^d} - {h_1},\;{h_1} \equiv {h_2} \equiv 1\;(\bmod \;2) $,}  \\
	Z_{2^d}           &\text{if ${h_2} \notin \{ {h_1}{,2^d} - {h_1}\} ,\;{h_1} \equiv {h_2} \equiv 1\;(\bmod \;2) $,}  \\
	\left\{ {2j|j \in Z_{2^{d - 1}}} \right\}  &\text{if ${h_1} \equiv {h_2} \equiv 2\;(\bmod \;4)  $.}
\end{cases}
$$

\noindent\textbf{Proof of Lemma}. For all $s,{v_1},{v_2} \in Z_{{2^{d - 1}}}$, we denote
$$s{\bar A^{(d)}}({v_1},{v_2}) = \left\{ {sb \bmod{2^d}|b \in {{\bar A}^{(d)}}({v_1},{v_2})} \right\}.$$

Let $t$ be an element from ${\bar A^{(d)}}({h_1},{h_2})$. Therefore,
$t = {h_1}{i_1} - {h_2}{i_2} \bmod{2^d}$ for some ${i_1},{i_2} \in Z_{{2^{d - 1}}}$.

Let ${h_i} \equiv 1\;(\bmod \;2)$ for some $i \in \{ 1,2\} $.  Without loss of generality, we suppose   ${h_1} \equiv 1\;(\bmod \;2)$. Then
$h_1^{ - 1}t = {i_1} - h_1^{ - 1}{h_2}{i_2} \mod{2^d}.$
So,
$t' = {i_1} - h \cdot {i_2} \mod{2^d},$
where $t' = h_1^{ - 1}t$, $h = h_1^{ - 1}{h_2}$.

Obviously,
${\bar A^{(d)}}({h_1},{h_2}) = {\bar A^{(d)}}({h_1},{h_1}h) = {h_1}{\bar A^{(d)}}(1,h).$

Now, we consider two cases.

{\bf Case 1}. Let $h$ be odd. For all  ${i_1},{i_2} \in {Z_{{2^{d - 1}}}}$, we have
$$
{i_1} - {i_2}h\not  \equiv
\begin{cases}
	{2^{d - 1}}\;(\bmod \;{2^d}) &\text{if $h = 1$,} \\
	{2^d} - 1\;(\bmod \;{2^d}) &\text{if $h = {2^d} - 1. $}
\end{cases}
$$
If $h \in \{ {3,5,7,...,2^d} - 3\} $, then
\begin{align*}
	{{\bar A}^{(d)}}(1,h) &= \bigcup\limits_{{j_2} = 0}^{{2^{d - 1}} - 1} {\left\{ {{j_1} - h \cdot {j_2}\;|{j_1} \in {Z_{{2^{d - 1}}}}} \right\}}=\\
	&={Z_{{2^{d - 1}}}} \cup \left\{ {{2^d} - h{{,2}^d} - h + {{1,\ldots,2}^{d - 1}} - h - 1} \right\} \cup\ldots\\
	&\cup \left\{ {{2^d} - 2h{{,2}^d} - 2h + {{1,...,2}^{d - 1}} - 2h - 1} \right\} \cup\\
	&\cup \left\{ {2h + {2^{d - 1}},2h + 1 + {2^{d - 1}},...,2h - 1} \right\}\cup\ldots\\
	&\cup \left\{ {h + {2^{d - 1}},h + 1 + {2^{d - 1}},...,h - 1} \right\} = Z_{2^d},
\end{align*}
where the operations addition and subtraction are over $\mathbb{Z}_{2^{d}}$.

Hence,
$${\bar A^{(d)}}(1,h) =
\begin{cases}
	{Z_{{2^d}}}\backslash \{ {2^{d - 1}}\} &\text{if $h = 1$,} \\
	{Z_{{2^d}}}\backslash \{ {2^d} - 1\} &\text{if $h = {2^d} - 1$,}  \\
	{Z_{{2^d}}} &\text{if $h \in \{ {3,5,...,2^d} - 3\} . $}
\end{cases}
$$

{\bf Case 2}. Let $h$ be even. From condition (\ref{E:5.18}), it follows that ${h_2} \equiv 2\;(\bmod \;4)$. Thus, $h \equiv 2\;(\bmod \;4)$. Hence,
\begin{align*}
	{{\bar A}^{(d)}}(1,h) &= \bigcup\limits_{{j_2} = 0}^{{2^{d - 1}} - 1} {\left\{ {{j_1} - h \cdot {j_2}|{j_1} \in {Z_{{2^{d - 1}}}}} \right\}}=\\
	&={Z_{{2^{d - 1}}}} \cup \left\{ {{2^d} - h{{,2}^d} - h + {{1,...,2}^{d - 1}} - h - 1\;} \right\} \cup\ldots\\
	&\cup \left\{ {2h,2h + 1,...,2h + {2^{d - 1}} - 1} \right\} \cup\\
	&\cup \left\{ {h + {2^{d - 1}},h + 1 + {2^{d - 1}}{{,...,2}^d} - {{2,2}^d} - 1,0,1,...,h - 1} \right\} = {Z_{{2^d}}}
\end{align*}
where the operations addition and subtraction are over $\mathbb{Z}_{2^{d}}$.

So, if ${h_i} \equiv 1\;(\bmod \;2)$ for some $i \in \{ 1,2\} $, then
$$
{\bar A^{(d)}}({h_1},{h_2}) =
\begin{cases}
	{Z_{{2^d}}}\backslash \{ {2^{d - 1}}\}, &\text{if $h_1 = h_2$,} \\
	{Z_{{2^d}}}\backslash \{ 2^d - h_1\}, &\text{if $h_2 = 2^d - h_1$,}  \\
	Z_{{2^d}}, &\text{if ${h_2} \notin \{ {h_1}{,2^d} - {h_1}\}.$}
\end{cases}
$$

Suppose ${h_1} \equiv {h_2} \equiv 2\;(\bmod \;4)$. Thus,
$t = 2\tilde t \; \bmod{{2}^d},$
where
$\tilde t = {\tilde h_1}{i_1} - {\tilde h_2}{i_2} \; \bmod{2^{d - 1}},$
${\tilde h_1} = {h_1}/2, \; {\tilde h_2} = {h_2}/2.$
Note that ${\tilde h_1} \equiv {\tilde h_2} \equiv 1\;(\bmod \;2)$.
From
$${Z_{{2^{d - 1}}}} = \left\{ {{{\tilde h}_1}{j_1} - {{\tilde h}_2}{j_2}\; \bmod{2^{d - 1}}|{j_1},{j_2} \in {Z_{{2^{d - 1}}}}} \right\},$$
we get
$${\bar A^{(d)}}({h_1},{h_2}) = \left\{ {2j|j \in {Z_{{2^{d - 1}}}}} \right\}.$$
\begin{flushright}
	End of Lemma proof. 
\end{flushright}

From Lemma and conditions (\ref{E:5.12}), (\ref{E:5.13}), it follows that we must consider four cases:
\begin{itemize}[noitemsep]
	\item {${r_1} \equiv {r_2} \equiv 1\;(\bmod \;2),$ }
	\item {${r_1} \equiv 1\;(\bmod \;2)$, ${r_2} \equiv 2\;(\bmod \;4)$,}
	\item {${r_1} \equiv 2\;(\bmod \;4)$, ${r_2} \equiv 1\;(\bmod \;2)$,}
	\item {${r_1} \equiv {r_2} \equiv 2\;(\bmod \;4)$.}
\end{itemize}

If ${r_1} \equiv {r_2} \equiv 1\;(\bmod \;2)$, then
\begin{equation}\label{E:5.32*}
	{r_1} \in \{ {q_2}{,2^{m - 1}} - {q_2}\}, \; {r_2} \in \{ {q_1}{,2^{m - 1}} - {q_1}\}.
\end{equation}
From condition (\ref{E:5.36}), we get
${r_1} \equiv {r_2} \equiv 3\;(\bmod \;4).$

For each $i,j \in \{ 1,2\} $, $i \ne j$, if ${r_j} = {2^{m - 1}} - {q_i}$, then ${q_i} \equiv 1\;(\bmod \;4)$ that contradicts (\ref{E:5.36}). Consequently,
${r_j} \ne {2^{m - 1}} - {q_i}$ for ${q_i} \equiv 1\;(\bmod \;4)$.
From Lemma and conditions (\ref{E:5.30}), (\ref{E:5.31}),  we get
\begin{equation}\label{E:5.33+}
	{b_1} + {c_1} \equiv 1\,(\bmod \,2), \; {b_2} + {c_2} \equiv 1\,(\bmod \,2).
\end{equation}

If $r_1 = q_2$, $r_2 = q_1$, then relations (\ref{E:5.12}), (\ref{E:5.13}) hold if and only if $c_1$, $c_2$, $b_1$, $b_2$ satisfy conditions
$${2^{m - 2}} \equiv {q_2}{2^{m - 2}} + {b_2} - {c_1}\;(\bmod \;{2^{m - 1}}), ~~~~
{2^{m - 2}} \equiv {q_1}{2^{m - 2}} + {b_1} - {c_2}\;(\bmod \;{2^{m - 1}}),$$
i.e.
\begin{equation}\label{E:5.33*}
	{c_1} = {b_2}, {c_2} = {b_1}.
\end{equation}
From (\ref{E:5.33+}) and (\ref{E:5.33*}), we get
${c_1} + {c_2} \equiv 1\,(\bmod \,2).$

Let $i,j \in \{ 1,2\} $, $i \ne j$. If ${r_j} \equiv 1\;(\bmod \;2),$ ${r_i} \equiv 2\;(\bmod \;4)$, then
\begin{equation}\label{E:5.33**}
	{r_j} \in \{ {q_i}{,2^{m - 1}} - {q_i}\}, ~~~ {r_i} \equiv {q_j} \equiv 2\;(\bmod \;4).
\end{equation}

From  (\ref{E:5.33**}), it follows that ${r_j} - 1\not  \equiv 1 - {q_j}\,(\bmod \;2)$. Therefore, from relations  (\ref{E:5.30}),  (\ref{E:5.31}) and Lemma, we get that condition (\ref{E:5.33**}) is impossible.

If ${r_1} \equiv {r_2} \equiv 2\;(\bmod \;4)$, then
${q_2}{2^{m - 2}} + {b_2} - {c_1}\; \equiv 1\;(\bmod \;2)$, ${q_1}{2^{m - 2}} + {b_1} - {c_2}\; \equiv 1\;(\bmod \;2).$
Thus,
\begin{equation}\label{E:5.34*}
	{b_1} + {c_2}\; \equiv 1\;(\bmod \;2),\; {b_2} + {c_1}\; \equiv 1\;(\bmod \;2).
\end{equation}

From Lemma and relations (\ref{E:5.30}), (\ref{E:5.31}), we have
${r_i} - 1 \in \{ 1 - {q_i}{,2^{m - 1}} - 1 + {q_i}\} \text{ for each } i \in \{ 1,2\},$
where
\begin{equation*}\label{E:5.35*}
	- {b_i} - {c_i}=
	\begin{cases}
		2^{m - 2} &\text{if ${r_i} - 1 = 1 - {q_i}$,} \\
		1 - {q_i} &\text{if ${r_i} - 1 = {2^{m - 1}} - 1 + {q_i}, $}
	\end{cases}
\end{equation*}
where the operations addition and subtraction are over ${{\mathbb{Z}}_{{2^{m - 1}}}}$.

If ${r_i} - 1 = 1 - {q_i}$ for some $j \in \{ 1,2\} $, then ${r_j} = 2 - {q_j}$. Hence, ${q_j} \equiv 0\;(\bmod \;4)$ that contradicts (\ref{E:5.18}).
So, there is only one relation ${r_i} - 1 = {2^{m - 1}} - 1 + {q_i}\;(\bmod \;{2^{m - 1}}) \text{ for each } i \in \{ 1,2\}.$
Thus,
\begin{equation}\label{E:5.36*}
	{r_i} = {q_i} \text{ for each } i \in \{ 1,2\}.
\end{equation}

If ${r_1} \equiv {r_2} \equiv 2\;(\bmod \;4)$, then $\pi $ is a permutation if and only if conditions (\ref{E:5.34*}), (\ref{E:5.36*}) hold and
${q_i} - 1 = {b_i} + {c_i} \bmod{2^{m - 1}} \text{ for each } i \in \{ 1,2\}.$
\begin{flushright}
	End of Theorem proof.
\end{flushright}

Let ${\rm{OM}}{{\rm{D}}_m}$ be the subset of ${\rm{M}}{{\rm{D}}_m}$ consisting of all orthomorphisms.
From Theorem, it follows that $\left| {{\rm{OM}}{{\rm{D}}_4}} \right| = {2^8}.$

Full and complete solutions for this problem were proposed by four team. The best one was given by the team of Jeremy Jean and Hugues Randriam (France).

\subsection{Problem ``JPEG Encoding''}
\hypertarget{pr-jpeg}{}
\subsubsection{Formulation}

In order to decrease the readability of the exchanged messages, Alice and Bob decided to encode their messages using JPEG image compression. They write (or draw) their message in a graphics software, save it as a JPEG file and then encrypt the resulting file using some encryption algorithm.

Let us describe the details of the JPEG encoding. The matrix of pixels is first divided into $8\times 8$ matrices, and then the matrices of the type presented below are obtained from them using discrete cosine transform (DCT) and quantization. An interesting characteristic of these matrices is that most of the non-zero data is concentrated in the upper left corner of the matrix, and most of the data in the lower right corner is 0. After that, the matrix is encoded using $0$'s and $1$'s.

\medskip
\noindent{\bf One example of the matrix encoding} is the following algorithm:
\begin{itemize}[noitemsep]
	\item[{\bf 1.}]{First, the zigzag rule is used to convert the $8\times 8$ matrix into a one-dimensional vector;}
	\item[{\bf 2.}]{Then the Exp-Golomb code is used to encode each number in the vector. Each number (aside from $0$, which is encoded as just one bit $0$) is encoded by three parts:
		\begin{itemize}[noitemsep]
			\item{{\it length}: a sequence of $1$'s corresponding to the length of the binary representation of the number, followed by $0$ to mark the end of the length sequence;}
			\item{{\it sign}: a bit representing the sign of the number: $0$ for negative, $1$ for positive number;}
			\item{{\it residual}: the binary representation of the number, with the leading $1$ omitted.}
		\end{itemize}
		For example, the number $47$ is encoded as the sequence $\underbrace{1111110}_{length}\underbrace{1}_{sign} \underbrace{01111}_{residual}$;}
	\item[{\bf 3.}]{All encoded sequences are then concatenated and a $6$-bit sequence is added to the front. These $6$ bits represent the number of non-zero elements in the encoded sequence.}
\end{itemize}

\begin{figure}[H]
	\includegraphics[width=\linewidth]{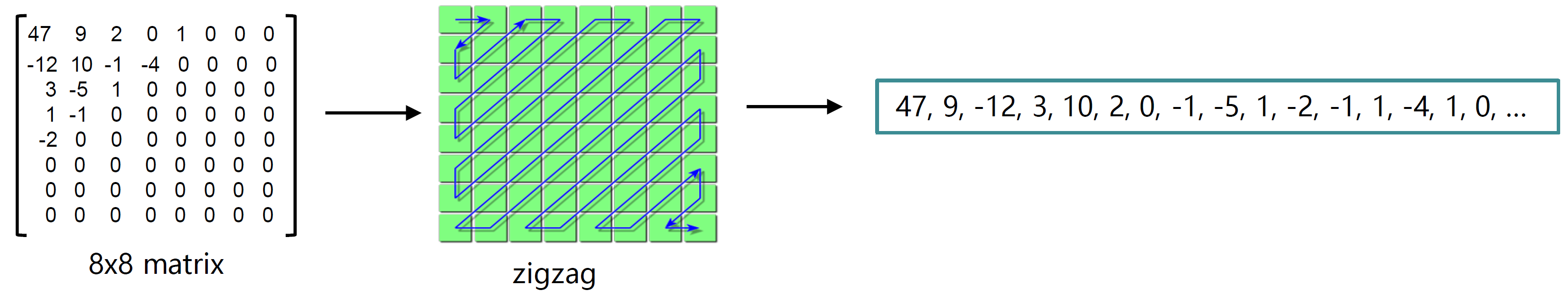}
	\vspace{-0.5cm}
	\caption{Zig-zag transformation of the matrix}
	\label{fig:zizzag}
\end{figure}

\noindent {\bf An example.} Let us consider how the algorithm works.
We can see that after Exp-Golomb coding (see Fig.~\ref{fig:zizzag}), the $8\times 8$ DCT quantized matrix above can be binarized using $91$ bits (see below). Note that using the inverse process of the encoding method, we can get the original $8\times 8$ matrix from these $91$ bits.
\begin{equation*}
	\resizebox{\textwidth}{!}{
		$\underbrace{001110}_{\#\text{ of non-zero elements}} \underbrace{1111110101111}_{47}\underbrace{111101001}_{9}\underbrace{111100100}_{-12}\underbrace{11011}_{3}\underbrace{111101010}_{10}\underbrace{11010}_{2}\underbrace{0}_{0}\underbrace{100}_{-1}\underbrace{1110001}_{-5}\underbrace{101}_{1}\underbrace{11000}_{-2}\underbrace{100}_{-1}\underbrace{101}_{1}\underbrace{1110000}_{-4}\underbrace{101}_{1}$
	}
\end{equation*}

\medskip
\underline{\bf Problem for a special prize!}
Your task is to design an encoding algorithm providing as short as possible output strings for the given 100\,000 matrices (\href{https://nsucrypto.nsu.ru/media/MediaFile/JPEG_Encoding-test_data_matrices.txt}{\textcolor{blue}{here}} is a file with  matrices, and non-zero elements of each matrix are concentrated in the upper left corner).
The less the sum of the lengths of the strings, the more scores you get for this problem. The encoding process must be reversible, that is, the original matrix can be obtained from the bit string using inverse coding.

\subsubsection{Solution}

By the authors opinion there were no great algorithms suggested. So, the problem remains open.

Let us discuss some criterions that were used for checking. An adequate algorithm for data processing should take into account the internal structure of the data involved. Therefore, the algorithms like: 1) get bits from the text file with matrices neglecting the matrix numeric data itself and compress them just as a stream of bits, scored low; 2) mechanical replacement of the suggested Exp-Golomb code with Huffman code or arithmetic code scored low; 3) the absence of the decoding procedure scored low; 4) not working code scored low.
The higher score got solutions which: 1) provided working encoder and decoder; 2) provided data analysis and were able to utilize the results of the data analysis in the algorithm; 3) provided good compression.

The initial authors' algorithm that used the Exp-Golomb code provides the compression size equal to 6\,694\,303 bits.
The lowest compression size 5\,878\,894 bits was achieved by team of Nhat Linh LE Tan and Viet Sang Nguyen	(France). Unfortunately, this algorithm just used the Huffman code instead of Exp-Golomb code.
Also, the team of Mikhail Kudinov, Alexey Zelenetskiy, and Denis Nabokov (Russia) suggested an interesting solution. They made some reasonable observations about the data and proposed changes into Exp-Golomb encoding depending on the position in the matrix which allows to improve compression. Their result was 5\,684\,601 bits. Unfortunately, there were some problems with executing the codes provided during the Olympiad.

\subsection{Problem ``Collisions''}
\hypertarget{pr-collisions}{}
\subsubsection{Formulation}

Consider a hash function $H$ that takes as its input a message $m$ consisting of $k\cdot n$ bits and returns an $n$-bit hash value $H(m)$. The message $m$ is at least one block long ($k \geqslant 1$), and can be split into $k$ blocks of $n$ bits each: $m_1$, $m_2$, $\ldots$, $m_k$.  Let $f$ be a function which takes an $n$-bit input and returns an $n$-bit output. 
We will use $\oplus$ to denote the bitwise exclusive-or operator.

\begin{figure}[H]
	\center
	\includegraphics[width=0.7\linewidth]{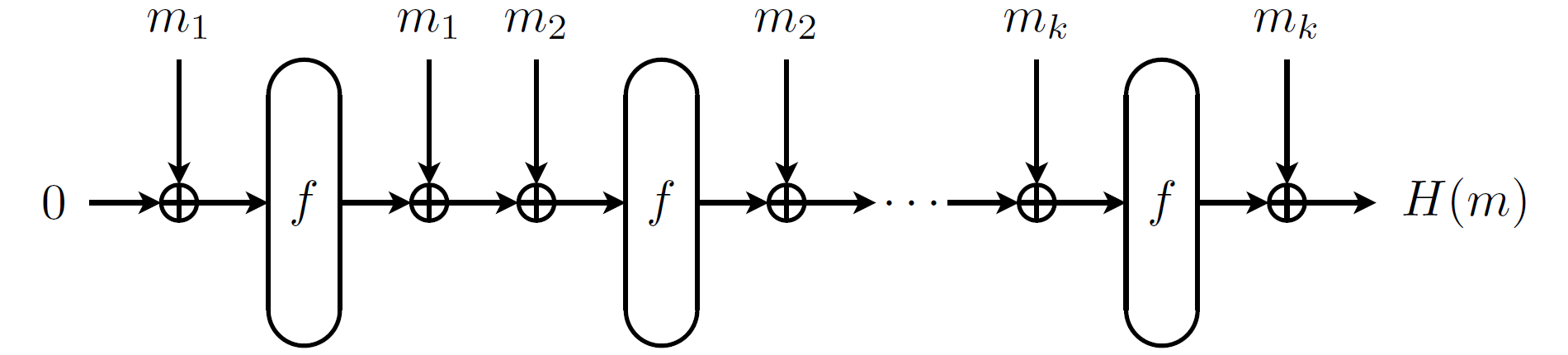}
	\caption{The hash function $H$.}
	\label{fig:h}
\end{figure}

The hash function $H$ is defined iteratively as follows:
\begin{equation*}
	h_i := m_i \oplus f(h_{i-1} \oplus m_i),
\end{equation*}
where all $n$ bits of $h_0$ are zero, and $H(m) := h_k$. An illustration of function $H$ is given in Fig.~\ref{fig:h}.

A \emph{collision} for $H$ is defined as a pair of distinct messages $(m,m')$ so that $H(m)$ = $H(m')$.
Given a message $m$ and its corresponding hash value $H(m)$, a \emph{second preimage} for $H$ is defined as a message $m' \neq m$ so that $H(m) = H(m')$.

Suppose that $f$ is a secret random function and that you have obtained $10\cdot n$ random different pairs $(x,f(x))$ of argument and value of the function $f$. Under these restrictions, solve the following problems. Algorithms in Q1 and Q2 must give a solution with a high probability ($>1/2$).

\begin{enumerate}
	\item[{\bf Q1}] {Propose an algorithm which finds a collision for $H$.}
	\item[{\bf Q2}] {Propose an algorithm which, given a message $m$ and its corresponding hash value $H(m)$, finds a second preimage $m'$ for $H$.}
	\item[{\bf Q3}] {Suppose that $n = 256$ bits and the message $m$ is ``{\tt A random matrix is likely decent}''. Find a second preimage $m'$ for this message.
		
		\noindent{\bf Remark 1.} The text message is converted into a bit sequence as follows: first, each character is converted into a $8$-bit integer according to the UTF-8 encoding, and then these integers are concatenated together using the big-endian ordering. For example, the string ``{\tt Hello}'' is converted into the sequence of integers $(72,101,108,108,111)$ which then gives the following binary string:
		0100100001100101011011000110110001101111. You can give your answer to this task in the form of a binary sequence or a hexadecimal sequence.
		
		\noindent {\bf Remark 2.} You can evaluate the hash function $H$ on any input message \href{https://nsucrypto.nsu.ru/archive/2020/round/2/task/8}{\textcolor{blue}{here}} \cite{nsucrypto-hash1}. The message being hashed should be presented as either a binary sequence or a hexadecimal sequence, starting with a symbol \texttt{b} or \texttt{h} which specifies the representation. \href{https://nsucrypto.nsu.ru/media/MediaFile/Collisions-Values_of_F.txt}{\textcolor{blue}{Here}} \cite{nsucrypto-hash2}
		you can find a list of values of $f$ on $512$ different inputs (binary sequences are presented as integers).
	}
\end{enumerate}

\subsubsection{Solution}

Let $||$ denote the concatenation of bit strings. Below we give solutions for all subproblems.

\textbf{Q1}.  It is easy to notice that $H(x||f(x)) = 0$ for any $n$-bit string. Therefore, for any two vectors $x,y$ with known values $f(x), f(y)$, messages $x||f(x)$ and $y||f(y)$ produce the same hash value $0$.

\textbf{Q2}. By {\bf Q1}, we can see that for any message $m$ and any $n$-bit string $x$ it holds $$H(x||f(x)||m) = H(m).$$
So, we can easily construct $10\cdot n$ preimages for any given message $m$.
Alternatively, one can append messages to the end: $H\big(m||H(m)\oplus x||H(m)\oplus f(x)\big) = H(m).$

\textbf{Q3}. This subproblem essentially asks one to apply their solution for Q2 to a specific example. The easiest solution is to append the string $0||f(0)$ to the message. The hexadecimal representation of the given message ``{\tt A random matrix is likely decent}'' is $$m={\tt 412072616e646f6d206d6174726978206973206c696b656c7920646563656e74}.$$
Taking the value of $f(0)$ from the given list, one can construct the following collision:
\begin{align*}
	m' &= 0||f(0)||m =\\
	& {\tt 0000000000000000000000000000000000000000000000000000000000000000}\\
	& {\tt ff1282609f458d732888e2736fd1b98cc36f809b1c116e77015b8d7d4d8996ae}\\
	& {\tt 412072616e646f6d206d6174726978206973206c696b656c7920646563656e74}.
\end{align*}

Let us describe also an \textbf{alternative solution for Q2} that was found by Andy Yu (Taiwan).
Let us denote $g_i = h_{i-1} \oplus m_i$ for $i=1,2,\ldots, k$. We then claim that
$$
h_j = \bigoplus_{i = 1}^{j} g_i \oplus f(g_i)
$$
for any $j=1,2,\ldots, k$.
The proof is by induction. Since $g_1 = h_0 \oplus m_1 = m_1$, we have
$h_1 = m_1 \oplus f(m_1) = g_1 \oplus f(g_1).$
Let $j>1$ and assume that $h_{j-1} = \bigoplus_{i = 1}^{j-1} (g_i \oplus f(g_i))$. Then
$$
h_j = m_j \oplus f(m_j \oplus h_{j-1}) = g_j \oplus h_{j-1} \oplus f(g_j) =
g_j \oplus f(g_j) \oplus \bigoplus_{i = 1}^{j-1} g_i \oplus f(g_i) = \bigoplus_{i = 1}^{j} g_i \oplus f(g_i),
$$
which proves the claim.
Note now that $H(m) = h_k = \bigoplus_{i = 1}^{k} g_i \oplus f(g_i).$ If we find a set of values $g'_1,g'_2,\ldots, g'_s$ such that $H(m) = \bigoplus_{i = 1}^{s} g'_i \oplus f(g'_i)$, we can easily construct a second preimage $m'$ by flipping the definition of $g_i$'s:
\begin{equation}\label{eq:NikkySol1}
	m'_j = g'_j \oplus h_{j-1} = g'_j \oplus \bigoplus_{i = 1}^{j-1} g'_i \oplus f(g'_i), \,\, j = 1,2,\ldots s.
\end{equation}

So, the task becomes the following: given the set of $10\cdot n$ pairs $\{(x_i, f(x_i))\}_{i=1}^{10\cdot n}$, find a subset of indices $i_1,\ldots,i_s$ such that
$H(m) = x_{i_1} \oplus f(x_{i_1}) \oplus \ldots \oplus x_{i_s} \oplus f(x_{i_s}).$
Let us denote $y_i = x_i \oplus f(x_i),\ i = 1,\ldots, 10\cdot n$. Then our goal is to express $H(m)$ as a linear combination of vectors $y_i$. Representing $y_i$'s as binary vectors of length $n$, we can easily solve this task by writing out and solving a system of binary linear equations with $n$ equations and $10\cdot n$ variables. But this works only if the value $H(m)$ is in the linear span of the vectors $y_i$. The probability of this event can be estimated as follows:
\begin{align*}
	&{\rm Pr}[H(m) \text{ is in the span of } y_i\text{'s}] \geqslant {\rm Pr}[y_i\text{'s span the whole space }\mathbb{F}_2^n] =\\
	= &\  {\rm Pr}[\text{Random binary } n\times 10\cdot n \text{ matrix has full rank } n] =\\
	= &\  \frac{(2^{10n}-1)(2^{10n}-2)(2^{10n}-4)\ldots(2^{10n}-2^{n-1})}{2^{10n^2}} = \prod_{i=0}^{n-1} (1 - 2^{-10n + i}) \geqslant\\
	\geqslant &\ 1 - \sum_{i=0}^{n-1} 2^{-10n + i} = 1 - 2^{-10n}(2^n - 1) \geqslant 1 - 2^{-9n}.
\end{align*}
Here the 4th line is obtained from the 3rd by repeatedly applying $(1-a)(1-b) \geqslant 1-a-b$.

\medskip
\noindent So, the algorithm is then the following:
\begin{itemize}[noitemsep]
	\item[{\bf 1.}] Calculate $y_i = x_i \oplus f(x_i)$ for $i=1,2\ldots 10\cdot n$.
	\item[{\bf 2.}] Construct an $n\times 10\cdot n$ matrix $A$ using $y_i$'s as its columns.
	\item[{\bf 3.}] Solve the linear system $A\cdot z = H(m)$. The probability of success of this step is at least $1 - 2^{-9n}$.
	\item[{\bf 4.}] Taking vectors $y_i$ for which $z_i = 1$, reconstruct the second preimage $m'$ using (\ref{eq:NikkySol1}). If $m' = m$, shuffle the order of $y_i$'s.
\end{itemize}

As well as the solution described above, notable solutions with extensive research was given by the team of Nhat Linh LE Tan and Viet Sang Nguyen (France), the team of Mircea-Costin Preoteasa, Gabriel Tulba-Lecu, and Ioan Dragomir (Romania).

\subsection{Problem ``Bases''}\label{bases}
\hypertarget{pr-bases}{}
\subsubsection{Formulation}

\underline{\bf Problem for a special prize!}
Let us consider the vector space $\mathbb{F}_2^r$ consisting of all binary vectors of length $r$. For any $d$ vectors $x^{i} = (x^i_1,\ldots,x^i_r)$, $i=1,\ldots,d$, $d>0$, it is defined the  componentwise product of these vectors equal to  $(x^1_1\ldots x^d_1,\ldots,x^1_r\ldots x^d_r)$. The empty product (when no element is involved in it) equals the all-ones vector.

Let $s\geqslant d>1$ be positive integers and let $r$ be defined by the formula $r = \sum_{i=0}^d {s\choose i}$, where ${s\choose i}$ denotes the binomial coefficient.
Let $\mathcal{B}$ be a basis of the vector space $\mathbb{F}_2^r$, and let $\mathcal{F}\subseteq\mathbb{F}_2^r$ be a family of $s$ binary vectors such that all possible componentwise products of up to $d$ vectors from the family $\mathcal{F}$ (including the empty product) form the basis $\mathcal{B}$.

Given $s,d,r$ defined above, describe all (or at least some) bases $\mathcal{B}$ for which such family~$\mathcal{F}$ exists or prove that such bases do not exist.

Suggest practical applications of such bases.

\medskip
\noindent {\bf Example.} Let $s = 2$, $d = 2$ and $r = 4$.
Consider the following family of $2$ vectors $\mathcal{F} = \{(1100),(0110)\}$.
Then all componentwise products of $0$, $1$ and $2$ vectors from the family $\mathcal{F}$ form the basis $\mathcal{B} = \{(1111),(1100),(0110),(0100)\}$ of  $\mathbb{F}_2^4$.

\subsubsection{Solution}

The problem ``determine what are the bases'' was not solved. This problem remains open.
The sub-problem  ``determine some bases'' was solved constructively by the team of  Mikhail Kudinov, Alexey Zelenetskiy, and Denis Nabokov (Russia). Let us describe the main ideas of this solution.

We will prove that such bases exist for all $s\geqslant d>1$ and give a construction of such bases.

Let ${\bf 1}$ be all-one vector and $r = \sum_{i=0}^d {s\choose i}$.
Suppose that there exists $\mathcal{F} \subseteq\mathbb{F}_2^r$ such that $\mathcal{F} = \{v_1,v_2,\ldots,v_s\}$ and $\mathcal{B} = \{v_{i_1}\ldots v_{i_k}\ |\ 1\leqslant i_1 < i_2 < \ldots < i_k \leqslant s \text{ and } 0\leqslant k \leqslant d\}$ is a basis of $\mathbb{F}_2^r$. Let $A$ be $(r\times r)$-matrix over $\mathbb{F}_2^n$ whose rows are exactly the vectors from $\mathcal{B}$. The rank of $A$ is equal to~$r$ since $\mathcal{B}$ is a basis.
Let $A^{(i)}$ denote the $i$-th column of $A$.  We number the rows of $A$ and, accordingly, the coordinates of $A^{(i)}$ as follows. The row corresponding to the vector $v_{i_1}v_{i_2}\ldots v_{i_k}$ we number as $i_1 i_2, \ldots,i_k$, the first row of $A$ we number as 0. For each $A^{(i)}$, the coordinate number 0 is nonzero and the coordinates $1, 2, \ldots, s$ determine the rest coordinates. Namely, the coordinate $i_1i_2\ldots i_k$ is equal to the product of
coordinates numbered $i_1, i_2,\ldots, i_k$.

{\bf Case $s = d$.} In this case $r = \sum_{i-0}^d {d\choose i} = 2^d$.
Let $x = (x_0,x_1,\ldots, x_{r-1})\in \mathbb{F}_2^r$ with $x_0=1$ and $x_1,\ldots,x_d$ determine $x_{d+1},\ldots,x_{r-1}$. The number of such vectors is equal to $2^d = r$. Only these vectors can be the columns of the matrix $A$. Since $A$ has $r$ columns and its rank is $r$, then $A$ (and as a consequence, a basis in $\mathbb{F}_2^r$) is uniquely defined by these vectors up to permutation of columns. Thus, if there are bases in $\mathbb{F}_2^r$, then the number of them is $r! = (2^d)!$.

Let us prove that these bases exist for an arbitrary $d$. Let us consider $\mathbb{F}_2^r$, $r = 2^d$, as a set of values vectors of all Boolean functions in $d$ variables. Since each Boolean function has the unique algebraic normal form (ANF), then the values vectors of all $2^d$ elementary monomial functions
$$\{1,\ x_1,\ x_2,\ \ldots,\ x_d,\ x_1x_2,\ \ldots,\ x_{d-1}x_d,\ \ldots,\ x_1\ldots x_d\}$$ form a basis in $\mathbb{F}_2^r$.

{\bf Case $s > d$.} Let us construct an invertible matrix $A$ (and as a consequence, a basis in $\mathbb{F}_2^r$) for an arbitrary $s>d$. Let the first column of $A$ be the vector $(1,0,0,\ldots,0)$. The next $s$ columns are
$$(1,1,0,\ldots,0), (1,0,1,\ldots,0),\ldots, (1,0,\ldots0,1,0,\ldots,0)$$.
We denote them as $A_1$. The next ${s\choose 2}$ vectors we denote as $A_2$. Each vector in $A_2$ has only four nonzero coordinate numbered $0,i,j,ij$, $1\leqslant i< j \leqslant s$. Analogically, the set $A_j$ consists of ${s\choose j}$ vectors and each vector has $2^j$ nonzero coordinates numbered  $0, i_1, i_2, \ldots, i_j , i_1i_2, i_1i_3, \ldots, i_1i_2\ldots i_j$, $1\leqslant i_1< i_2 < \ldots, i_j \leqslant s$.

The matrix $A$ constructed above is a triangular matrix and each element on the main diagonal is equal to 1. Therefore, the matrix $A$ is invertible.  Any permutation of the columns gives us a new matrix, whose rows give us a basis. Thus, we have $\geqslant r!$ bases in $\mathbb{F}_2^r$.

\subsection{Problem ``AES-GCM''}
\hypertarget{pr-aes}{}
\subsubsection{Formulation}

Alice is a student majoring in cryptography. She wants to use AES-GCM-256 to encrypt the communication messages between her and Bob (for more details of GCM, we refer to \cite{07-Dworkin}). The message format is as follows:

\medskip

\begin{footnotesize}
\begin{tikzpicture}[x=0.61pt,y=0.75pt,yscale=-1,xscale=0.9]
	\draw  [fill={rgb, 255:red, 255; green, 255; blue, 0 }  ,fill opacity=1 ] (100,100) -- (180,100) -- (180,130) -- (100,130) -- cycle ;
	\node at (140,115) {Header};
	\node at (140,145) {8 bytes};
	
	\draw  [fill={rgb, 255:red, 0; green, 255; blue, 0 }  ,fill opacity=1 ] (180,100) -- (340,100) -- (340,130) -- (180,130) -- cycle ;
	\node at (260,115) {Initialization Vector};
	\node at (260,145) {12 bytes};
	
	\draw  [fill={rgb, 255:red, 75; green, 150; blue, 255 }  ,fill opacity=1 ] (340,100) -- (580,100) -- (580,130) -- (340,130) -- cycle ;
	\node at (460,116.5) {Encrypted Payload};
	\node at (460,145) {$n$ bytes};
	
	\draw  [fill={rgb, 255:red, 0; green, 220; blue, 0}  ,fill opacity=1 ] (580,100) -- (740,100) -- (740,130) -- (580,130) -- cycle ;
	\node at (660,116.5) {Authentication Tag};
	\node at (660,145) {16 bytes};
	
\end{tikzpicture}
\end{footnotesize}

\noindent However, Alice made some mistakes in the encryption process since she is new to AES-GCM.
Your task is to attack the communications.

\begin{enumerate}
	\item[{\bf Q1}] {
		You intercepted some messages sent by Alice. You can find them in the directory \href{https://nsucrypto.nsu.ru/media/MediaFile/AES-GCM-Task_1.7z}{\textcolor{blue}{``Task\_1''}}.
		Also, you know that the plaintext (unencrypted payload) of the first message (0.message) is
		``{\tt Hello, Bob! How's everything?}'' \newline(without quotes, encoded in UTF-8).
		Try to decrypt any message in the directory \href{https://nsucrypto.nsu.ru/media/MediaFile/AES-GCM-Task_1.7z}{\textcolor{blue}{``Task\_1''}}.
	}
	
	\item[{\bf Q2}]
	In this task, you further know that the AAD (additional authenticated data) used by Alice in each 	message is Header\,$||$\,Initialization Vector:
\medskip

	\begin{footnotesize}
	\begin{tikzpicture}[x=0.61pt,y=0.75pt,yscale=-1,xscale=0.9]
		\draw (100,130) -- (100,155);
		\draw  [fill={rgb, 255:red, 255; green, 255; blue, 0 }  ,fill opacity=1 ] (100,100) -- (180,100) -- (180,130) -- (100,130) -- cycle ;
		\node at (140,115) {Header};
		
		\draw  [fill={rgb, 255:red, 0; green, 255; blue, 0 }  ,fill opacity=1 ] (180,100) -- (340,100) -- (340,130) -- (180,130) -- cycle ;
		\node at (260,115) {Initialization Vector};
		\node at (220,145) {Additional Authenticated Data};
		\draw (340,130) -- (340,155);
		
		\draw  [fill={rgb, 255:red, 75; green, 150; blue, 255 }  ,fill opacity=1 ] (340,100) -- (510,100) -- (510,130) -- (340,130) -- cycle ;
		\node at (430,116.5) {Encrypted Payload};
		
		\draw  [fill={rgb, 255:red, 0; green, 220; blue, 0}  ,fill opacity=1 ] (510,100) -- (670,100) -- (670,130) -- (510,130) -- cycle ;
		\node at (590,116.5) {Authentication Tag};
	\end{tikzpicture}
	\end{footnotesize}

	You want to tamper some messages in the directory \href{https://nsucrypto.nsu.ru/media/MediaFile/AES-GCM-Task_2.7z}{\textcolor{blue}{``Task\_2''}}.
	You pass this task if you can modify at least one bit in some message so that Bob can still decrypt the message successfully.
	
	\item[{\bf Q3}]
	Alice has noticed that the messages sent by her have been tampered with. So she decides to enhance the security of her encryption process.
	Instead of using Header\,$||$\,Initialization Vector as the additional authenticated data (AAD), Alice further generates 8 bytes data $X$ by some deterministic function $f$ and the AES secret key $K$,\,where
	$$X = f(K).$$
	
	In each message, she uses Header\,$||$\,Initialization Vector\,$||$\,$X$ as the AAD.
	
	You also intercepted some messages sent by Alice, see these messages in the directory \href{https://nsucrypto.nsu.ru/media/MediaFile/AES-GCM-Task_3.7z}{\textcolor{blue}{``Task\_3''}}.
	Try to tamper any message!
	
	\item[{\bf Q4}] \underline{\bf Bonus problem (extra scores, a special prize!)}
	
	You have successfully tampered with the messages in Q2. However, the attacks will be easy to
	detect if the tampered message cannot be decrypted to some meaningful plaintext.
	
	In this task, try to tamper the messages in Q2 so that the tampered message can still be decrypted to some plaintext that people can understand. {\bf Remark:} Tampering with the Header or Initialization Vector of a message will not be accepted as a solution, you need to tamper with the encrypted payload to produce some other ciphertext which did not appear in any message included.
	
\end{enumerate}

\subsubsection{Solution}

Let us give solutions or ideas for all subproblems.

\textbf{Q1.} Note that blocks of the ciphertext $C_i$ are obtained by XORing blocks of the plaintext $P_i$ with the values $E_k(CB_i)$. The values $E_k(CB_i)$ depend on the $IV$ and some other parameters which are common for all messages within one subproblem. Going through the messages, we can see that the messages number $0$, $5$ and $6$ all use the same initialization vector. Since we know the plaintext for the message number $0$, we can compute the first 29 bytes of the values $E_k(\cdot)$ for this $IV$ and use them to decipher the entirety of the $20$-byte message number $5$ and $29$ symbols of the $46$-byte message number $6$:
\begin{align*}
	& m_5 = \texttt{Lincoln Park, 10:15.} \\
	& m_6 = \texttt{Nostalgia is a eternal motif}
\end{align*}

\textbf{Q2.} In this subproblem, the messages number 1 and 6 also have the same initialization vector. We can apply the \textbf{Forbidden Attack} \cite{FA} to reconstruct the secret value $H$, which will allow us to forge messages by changing the ciphertext and recalculating the Authentication Tag. In this solution, we will briefly describe the attack.

Let $A = A_1 || A_2 || \ldots || A_m$ be the AAD of a message, and let $C = C_1 || C_2 || \ldots || C_n$ be the encrypted payload. Then the Authentication Tag can be presented as follows:
\begin{equation}\label{eq:HuaweiAES1}
	{\rm AuthTag} = E_k(CB_0) \oplus \sum_{i=1}^{m+n+1} T_i H^{m+n+2-i},
\end{equation}
where $T = A_1 || A_2 || \ldots || A_m || C_1 || C_2 || \ldots || C_n || (len(A) || len(C))$ and all operations are performed in the Galois field $\mathbb{F}_{2^{128}}$.

Let us consider (\ref{eq:HuaweiAES1}) as an equation which we want to solve for $H$. Since we know the AuthTag, the AAD and the ciphertext for every message, each coefficient in this equation is known except for $E_k(CB_0)$. However, since the messages number 1 and 6 have the same $IV$, they also have the same value $E_k(CB_0)$. Subtracting equations of the form (\ref{eq:HuaweiAES1}) constructed for the messages number 1 and 6 one from another, we obtain the following equation:
\begin{equation*}
	{\rm AuthTag}_1 - {\rm AuthTag}_6 = g(H),
\end{equation*}
where $g(H)$ is a polynomial in the variable $H$ with all coefficients known. We can find the root of it in the field $\mathbb{F}_{2^{128}}$:
\begin{align*}
	H =&\ a^{126} + a^{125} + a^{122} + a^{120} + a^{119} + a^{116} + a^{114} + a^{111} +
	      a^{110} + a^{107} + a^{99} +  a^{96} + a^{95} + a^{94} \\
+&\  a^{93} + a^{92}  + a^{90} + a^{89} + a^{87} + a^{85} + a^{84} + a^{83} + a^{82} + a^{81} + a^{80} + a^{78} + a^{76} + a^{73} + a^{67}  \\
+&\ a^{66} + a^{62} + a^{61} + a^{60} + a^{59} + a^{56} +
	a^{53} + a^{52} + a^{49} + a^{47} + a^{45} + a^{40} + a^{39} + a^{38} + a^{37} \\
+&\ a^{36} + a^{35} + a^{34} + a^{33} + a^{29} + a^{28} + a^{24} + a^{22}  + a^{21} +	a^{19} + a^{18} + a^{17} + a^{16} + a^{14} + a^{11} \\
+&\ a^{10} + a^9 + a^6 + a^4 + a^2,
\end{align*}
where $a$ is the generator of the field.
Knowing $H$, we can easily find $E_k(CB_0)$ and calculate the Authentication Tag for any ciphertext which was obtained using the same $IV$ as in the messages number 1 and number 6.

\textbf{Q3.} Observing messages from the subproblem, we can notice that the messages number 1, 3 and 7 have the same Header $h$, the same $IV$ and the same length of the ciphertext $len(C^j)$, $j = 1,3,7$. Let us split the Initialization Vector $IV = IV_0 || IV_1$ so that the AAD for each of the three messages can be written as $A = A_1 || A_2$, where $A_1 = h || IV_0$ and $A_2 = IV_1 || X || 0^{32}$. Then for $j=1,3,7$, we have:
$$
{\rm AuthTag}_j = E_k(CB_0) \oplus A_1 H^{23} \oplus A_2 H^{22} \oplus C_1^j H^{21} \oplus C_2^j H^{20} \oplus	 \ldots\oplus C_{20}^j H^{2} \oplus (len(A) || len(C) H.
$$
Here, we do not know $E_k(CB_0)$ and we also do not know $A_2$ since it contains the secret value $X = f(K)$. However, since the degrees of all three equations are the same, when we subtract one from another, the term with $A_2$ vanishes along with $E_k(CB_0)$. So, we can still apply the method used in $Q2$ to solve these equations for $H$.
After trying all possible combinations, we find the only value of $H$ which satisfies all equations at once:
\begin{align*}
	H =&\ a^{123} + a^{122} + a^{112} + a^{110} + a^{107} + a^{102} + a^{100} + a^{99} +
a^{97} + a^{96} + a^{95} + a^{92} + a^{90} + a^{87} + a^{85} \\
+&\ a^{83}  + a^{82} + a^{81} + a^{78} + a^{77} + a^{74} + a^{73} + a^{71} + a^{70} + a^{65} + a^{63} +
a^{62} + a^{60} + a^{59} + a^{58} + a^{57} \\
+&\ a^{54} + a^{53} + a^{50}  + a^{49} +
a^{47} + a^{45} + a^{43} + a^{42} + a^{41} + a^{37} + a^{36} + a^{32} + a^{30} +
a^{28} + a^{23} + a^{13} \\
+&\ a^{12} + a^{10} + a^7 + a^5 + a^3 + 1.
\end{align*}
Knowing $H$, we can once again modify any of the ciphertexts of the messages number 1, 3 or 7 and recalculate the Authentication Tag.

\textbf{Q4.} This subproblem remains open in general since there were no complete theoretical solutions given. However, many different approaches were presented to modify these particular messages utilizing the properties of the natural language.

Some participants suggested that we can flip the least significant bits in parts of the ciphertext in order to obtain a text with a ``typo''. Alternatively, we can try shuffling parts of ciphertexts encrypted with the same $IV$, which may produce a readable text, although likely not semantically connected.

Other participants used the properties of the natural English language to decipher the messages number 1 and 6 by hand. Note that, since the messages use the same $IV$, if we XOR the shorter ciphertext $C^6$ with the part of the longer ciphertext $C^1$, we will get $$C^6 \oplus C^1 = P^6 \oplus P^1.$$
Trying to find pairs of texts $P^1, P^6$ that are readable and sum to $C^6 \oplus C^1$ by hand, it is possible to discover the following two texts:
\begin{align*}
	& P^6 =\ \texttt{``Do not you want to know who has taken it?'' cried his wife impatiently.} \\
	& P^1 =\ \texttt{However little known the feelings or views of such a man may be on his }
\end{align*}
Note that we cannot be completely sure that these texts were the original messages, and we also cannot guarantee which text is $P^1$ and which is $P^6$.
However, it is highly likely we correctly decrypted the message number 6. We can now replace it with an arbitrary new message $\tilde{P}^6$ of the same length, and its corresponding ciphertext can be calculated as follows: $\tilde{C}^6 = \tilde{P}^6 \oplus C^6 \oplus P^6$. We are also able to calculate an Authentication Tag for this new message as we have solved Q2 and know $H$.

The most complete solutions to this problem were given by the team of Himanshu Sheoran, Sahil Jain, and Tirthankar Adhikari (India), the team of Mikhail Kudinov, Alexey Zelenetskiy, and Denis Nabokov	 (Russia), the team of Pham Cong Bach, Phu Nghia Nguyen, and Ngan Nguyen	(Vietnam), the team of Roman Sychev, Diana Bespechnaya, and Nikolay Prudkovskiy (Russia), the team of Roman Lebedev, Vladimir Sitnov, Ilia Koriakin (Russia).

\bigskip
\noindent{\bf Acknowledgments.}
We thank Alexey Oblaukhov for valuable comments and fruitful discussions.

\end{document}